\documentclass{aa}  

\usepackage{color}

\usepackage{graphicx}
\usepackage{txfonts}

\usepackage{amssymb}
\usepackage{multirow}
\usepackage{pifont}
\usepackage{diagbox}

\makeatletter
    \newcommand{\rmnum}[1]{\romannumeral #1}
    \newcommand{\Rmnum}[1]{\expandafter\@slowromancap\romannumeral #1@}
\makeatother
\begin{document}

      %\title{Number asymmetry between the L4 and L5 Jupiter Trojan swarms driven by the Jumping-Jupiter migration}   
      
      \title{Asymmetry in the number of L4 and L5 Jupiter Trojans driven by jumping Jupiter}

%   \subtitle{I. Overviewing the $\kappa$-mechanism}

     \author{Jian Li
         \inst{1,2},
          Zhihong Jeff Xia\inst{3} 
          \and
          Fumi Yoshida\inst{4,5}
          \and
          Nikolaos Georgakarakos\inst{6,7}
          \and
          Xin Li\inst{8}
          }

   \institute{School of Astronomy and Space Science, Nanjing University, 163 Xianlin Avenue, Nanjing 210023, PR China\\
              \email{ljian@nju.edu.cn}
         \and
         Key Laboratory of Modern Astronomy and Astrophysics in Ministry of Education, Nanjing University, Nanjing 210023, PR China
        \and
             Department of Mathematics, Northwestern University, 2033 Sheridan Road, Evanston, IL  60208, USA
         \and
         University of Occupational and Environmental Health, 1-1 Iseigaoka, Yahata, Kitakyusyu 807-8555, Japan
         \and
         Planetary Exploration Research Center, Chiba Institute of Technology, 2-17-1 Tsudanuma, Narashino, Chiba 275-0016, Japan 
         \and New York University Abu Dhabi, PO Box 129188 Abu Dhabi, United Arab Emirates
         \and Center for Astro, Particle and Planetary Physics (CAP$^3$), New York University Abu Dhabi, PO Box 129188 Abu Dhabi, United Arab Emirates 
         \and Department of Statistics and Data Science, Southern University of Science and
         Technology of China. No 1088, Xueyuan Rd., Xili,\\ Nanshan District, Shenzhen, Guangdong, 518055, PR China\\
             }
   \date{Received September 15, 1996; accepted March 16, 1997}

% \abstract{}{}{}{}{}  
% 5 {} token are mandatory
 
  \abstract
   {More than 10000 Jupiter Trojans have been detected so far. They are moving around the L4 and L5 triangular Lagrangian points of the Sun-Jupiter system and their distributions can provide important clues to the early evolution of the Solar System.}
   {The number asymmetry of the L4 and L5 Jupiter Trojans is a longstanding problem. We aim to test a new mechanism in order to explain this anomalous feature by invoking the jumping-Jupiter scenario.}
    {First, we introduce the orbital evolution of Jupiter caused by the giant planet instability in the early Solar System. In this scenario, Jupiter could undergo an outward migration at a very high speed. We then investigate how such a jump changes the numbers of the L4 ($N_4$) and L5 ($N_5$) Trojans.}
   {The outward migration of Jupiter can distort the co-orbital orbits near the Lagrangian points, resulting in L4 Trojans being more stable than the L5 ones. We find that, this mechanism could potentially explain the unbiased number asymmetry of $N_4/N_5\sim1.6$ for the known Jupiter Trojans. The uncertainties of the system parameters, e.g. Jupiter's eccentricity and inclination, the inclination distribution of Jupiter Trojans, are also taken into account and our results about the L4/L5 asymmetry have been further validated. However, the resonant amplitudes of the simulated Trojans are excited to higher values compared to the current population. A possible solution is that collisions among the Trojans may reduce their resonant amplitudes.}
   {}

   
   \titlerunning{Asymmetry in the number of L4 and L5 Jupiter Trojans}

   \maketitle
%
%-------------------------------------------------------------------

\section{Introduction}

%Thorough 

The Jupiter Trojans, also known as Jupiter’s co-orbitals, are small bodies that share the orbit of Jupiter. They essentially evolve around the L4 and L5 Lagrangian points, which respectively leads and trails Jupiter by an angular distance of $\sim60^{\circ}$. Considering the currently observed minor planets in the Solar System, the Jupiter Trojan population comprises over ten thousand bodies and is the second largest group, only smaller than the main belt asteroids. Besides the importance in number, the Jupiter Trojans also exhibit many peculiar physical and dynamical features, which have provided glimpses into the early history of the outer Solar System, e.g. the formation and evolution of the Jovian planets.

There are various studies on the origin of the Jupiter Trojans. Before the Nice model \citep{tsig05} was proposed, Jupiter Trojans  were thought to be a population of planetesimals near the Jupiter's orbit captured by Jupiter into the Trojan region during Jupiter's accumulation phase \citep{marz98a}.  After the Nice model, the most likely explanation is that the radial movement of giant planets scattered icy planetesimals in the outer Solar System, and some of these planetesimals from the present-day Kuiper belt that fell inward were captured into the Trojan region by Jupiter \citep{morb05, nesy12}.

%the most likely explanation is that the radial movement of giant planets scattered planetesimals in the giant planet region and the outer icy planetesimal belt (present-day Kuiper belt), and some of these planetesimals that fell inward were captured into the Trojan orbits by Jupiter \citep{morb05, nesy12}

Size frequency distributions of Jupiter Trojans larger than a few tens of kilometers have been reported by \cite{jewi00} , who detected Jupiter Trojans in data obtained from Kuiper belt object surveys, and by \cite{szab07}, who examined the SDSS moving object catalog.
On the other hand, the size frequency distribution of Jupiter Trojans has been estimated by \cite{yosh05}, \cite{yosh08}, \cite{wong15} \cite{yosh17}, \cite{ueha22} using Subaru + wide-field camera data. Thanks to their works, the size frequency distribution of Jupiter Trojans larger than 2 km in diameter has been known so far.
\cite{yosh08} and \cite{ueha22} compared the size frequency distributions of the L4 and L5 swarms. \cite{yosh08} pointed out that the L4 and L5 swarms have slightly different size frequency distribution from each other, using a few ten Jupiter Trojan samples. Later, this result was overturned by \cite{ueha22}, using larger sample of L5 Jupiter Trojans detected by a similar instrument but with a wider field of view camera. They concluded that the size frequency distribution of Jupiter Trojans larger than 2 km in diameter was very similar between the L4 and L5 swarms.

\cite{yosh19} and \cite{yosh20} also compared the size frequency distribution of Jupiter Trojans to that of main belt asteroids and Hilda asteroids and then showed that the size frequency distributions of Jupiter Trojans and Hildas are totally different from that of main belt asteroids. In addition, they compared the current size distribution of Jupiter Trojans with the size frequency distributions of the impactors that created the craters of Pluto and Charon \citep{sing19} and noted that their size distributions may be similar. This supports the idea, as proposed in the Nice model \citep{morb05}, that the present-day Jupiter Trojans are icy planetesimals that were scattered during planetary migration and captured by Jupiter into the Trojan region.

The composition and surface properties of the Jupiter Trojans are still poorly understood. What we do know so far are, 
(1) this population is more uniform than the main belt asteroids, 
(2) their albedos are low, 
(3) visible and near-infrared spectra are divided into two groups: red or neutral \citep{szab07, roig08, forn07}, and
(4) no clear feature is found in the visible-near red spectrum \citep{dott06, forn07, yang07, meli08}.
The bulk density of Jupiter Trojans is estimated to be $\sim 0.8-1.0$ g cm$^{-3}$ from the rotation period distribution of Jupiter Trojans with a diameter of 2 km $< D <$ 40 km and the binary Jupiter Trojan system, (617) Patroclus-Menoetius system \citep{chan21, marc06, muel10, buie15, bert20}.

%(For Fumi) Size distribution and the associated works: observation and origin, e.g. the difference in size distribution of the two swarms could cause a different probability of observation? The slope of the SFD for the two clouds match well (Kotomi Uehata et al. 2022, AJ)?

%(For Fumi) Inclination distribution: observation and origin (i.e. Nice model)

%(For Fumi) Other peculiar physical and dynamical features

Recently, we noticed the apsidal asymmetric-alignment of Jupiter Trojans \citep{li2021}. We have shown that the L4 and L5 swarms are both clustered in the longitude of perihelion ($\varpi$), around the locations that are $+60^{\circ}$ and $-60^{\circ}$ away from Jupiter's longitude of perihelion ($\varpi_{\mbox{\scriptsize{J}}}$) respectively. This peculiar architecture is caused by Jupiter's eccentricity of 0.05, as there exist two corresponding equilibrium points located at the positions $\varpi-\varpi_{\mbox{\scriptsize{J}}}=\pm60^{\circ}$ in the phase space of the dynamics of co-orbital motion. We further proved that the clustering of $\varpi$ of the Trojans is a natural consequence of the secular perturbation of an eccentric Jupiter, even when the initial distribution of $\varpi$ is uniform or a Gaussian. We then concluded that the apsidal asymmetric-alignment of Jupiter Trojans is robust and will persist for a considerably long timescale.

For the Jupiter Trojans, the number difference between the L4 and L5 swarms remains most mysterious. Ever since \citet{shoe89}, people have noticed that the L4 swarm is more populated. Nevertheless, at that time, this was because the L5 swarm was outshined by the Milky Way and was hard to observe. As more and more Jupiter Trojans have been discovered, the L4/L5 number asymmetry becomes a reality and is well accepted.

\subsection{L4/L5 number asymmetry}

\begin{figure}
 \hspace{0 cm}
  \includegraphics[width=8.5cm]{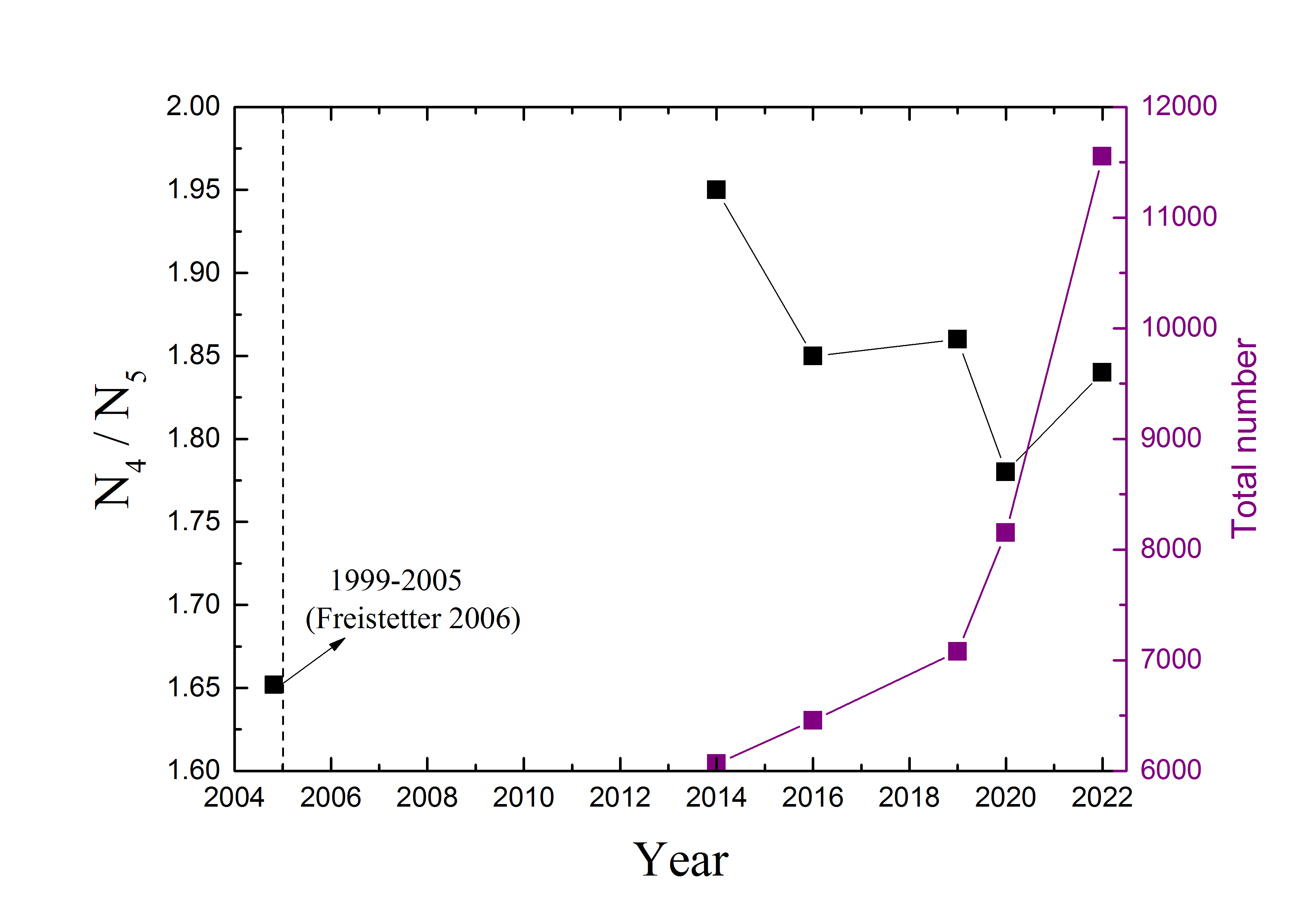}
  \caption{History of asymmetry in the number of the L4 and L5 Jupiter Trojan discoveries (black squares). The early data before 2005 is extracted from \citet{frei06}. As for the observed Jupiter Trojans after 2014, the asymmetry is obtained by using the MPC data recorded in our database; their total numbers at different epochs are also presented (purple squares).}
  \label{year}
\end{figure}

\begin{figure}
  \centering
  \begin{minipage}[c]{0.5\textwidth}
  \centering
  \hspace{0cm}
  \includegraphics[width=8.5cm]{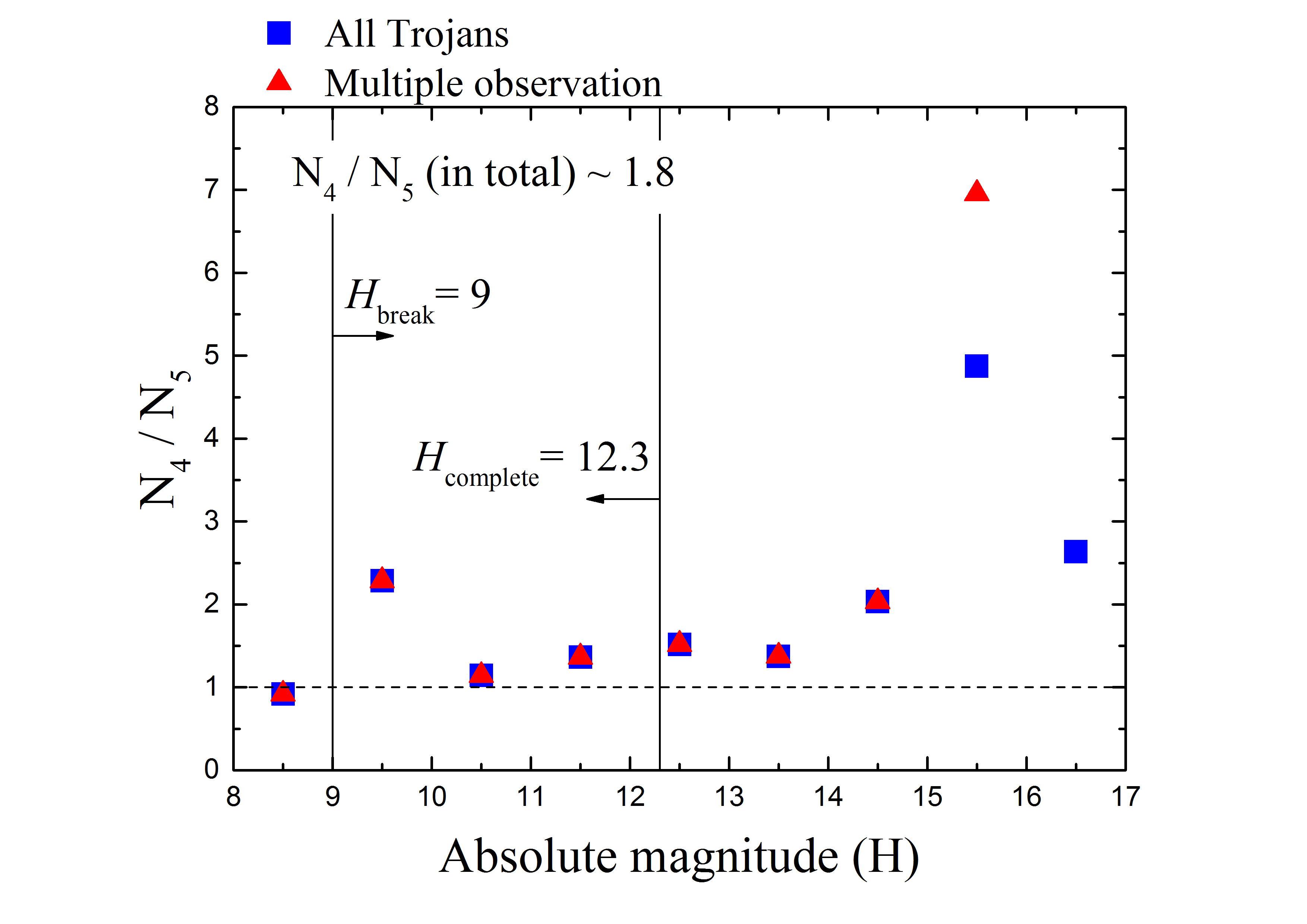}
  \end{minipage}
  \begin{minipage}[c]{0.5\textwidth}
  \centering
  \hspace{0cm}
  \includegraphics[width=8.5cm]{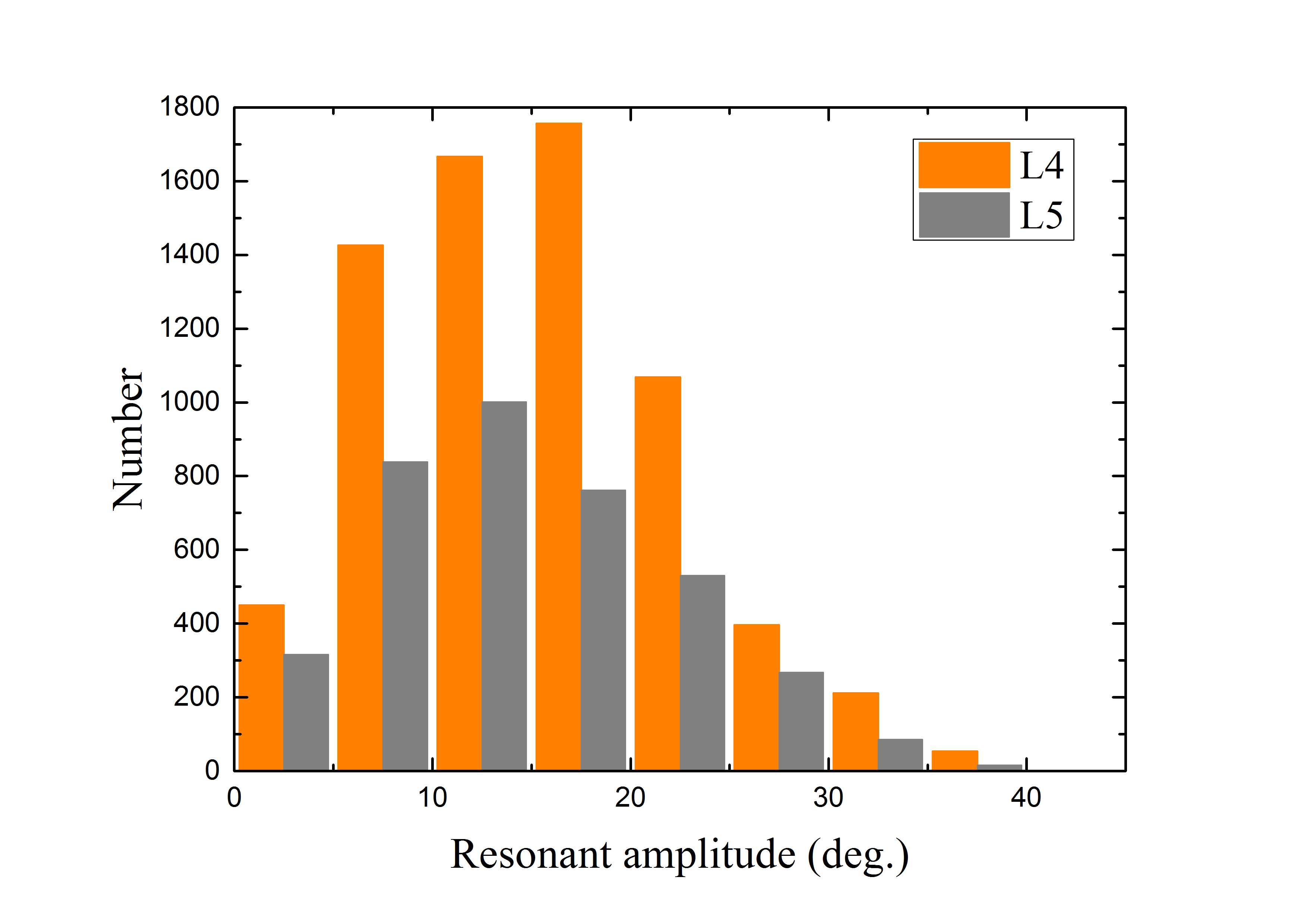}
  \end{minipage}
    \caption{For the latest observation of Jupiter Trojans taken from the MPC, at the epoch of 2022 January 26: (\textit{upper panel}) The L4/L5 number asymmetry at different absolute magnitudes ($H$). The blue square indicates the consideration of all Trojans, while the red triangle is only for the objects observed at multiple oppositions. The Trojans with $H\le H_{\mbox{\scriptsize{complete}}}=12.3$ are observationally complete and the ones with $H\ge H_{\mbox{\scriptsize{break}}}=9$ should have reached a state of collisional equilibrium \citep{li2018}. (\textit{lower panel})  Distribution of resonant amplitudes for the Trojans in the L4 (orange) and L5 (grey) swarms.}
  \label{RealJTs}
\end{figure}  

Fig. \ref{year} shows how the number difference between the L4 and L5 Trojan swarms has varied over the last two decades (indicated by the black symbols). In the years between 1999-2005, the leading-to-trailing number ratio $N_4/N_5$ was always around 1.67 \citep{frei06}. Later, \citet{szab07} analysed the distribution of about 860 Jupiter Trojans known at that time. By taking into account possible selection effects and the faint completeness limit, they determined a theoretical number ratio of $N_4/N_5=1.6\pm0.1$. As more and more Jupiter Trojans have been discovered in the past few years (see Fig. \ref{year}, purple symbols), reaching a considerably large number over 10000, we find that this ratio has increased to about 1.8.

As of 2022 January 26, 11552 Jupiter Trojans have been registered in the Minor Planet Center\footnote{https://minorplanetcenter.net/iau/lists/JupiterTrojans.html} (MPC). Fig. \ref{RealJTs}(a) shows that the number of L4 Trojans does exceed the number of L5 Trojans nearly in the entire range of the absolute magnitude ($H$) (equivalent to the size) and the overall leading-to-trailing number ratio is 1.84. The largest-sized Trojans with $H< H_{\mbox{\scriptsize{break}}}=9$ have not reached the state of collisional equilibrium \citep{morb09}, so the associated value of $N_4/N_5=0.92<1$ could be meaningless. Nevertheless, this population is rather small comprising only 24 objects and has little contribution to the overall number asymmetry. Regarding the smallest-sized Trojans with $H>15$, they are far beyond the observationally complete limit of $H_{\mbox{\scriptsize{complete}}}=12.3$. Currently only about 860 such faint objects have been observed, but their total number is estimated to be approximately half a million according to our previous work \citep{li2018}. This suggests that the intrinsic value of the overall leading-to-trailing number ratio is to be further improved by future surveys.  

%The samples with $H<12.3$ are believed to be observationally complete since no brighter objects have been discovered since mid-2006

% This population includes many objects that only have short-arc observations which may induce inaccurate orbit determinations for them.

Among the known Trojans, 10854 objects have been observed at multiple oppositions. Fig. \ref{RealJTs}(a) shows that at $H<15$, the $N_4/N_5$ distribution for that group (red triangles) is nearly the same as the one of the entire Trojan population (blue squares). This is easy to understand. First, because the multiple-opposition group represents a very large fraction ($\sim94\%$) of the Trojan population. Secondly, even if observations were made at only a single opposition, the obtained Lagrangian point classification (i.e. near L4 or L5) and absolute magnitude are quite confidential. Consequently, it is reasonable to consider only the multiple-opposition Trojans, whose orbital elements are well determined. Then we calculated their resonant amplitudes ($A$) through numerical integration, and the combined distribution for both the L4 (orange) and L5 (grey) swarms is plotted in Fig. \ref{RealJTs}(b). As well, it is obvious that a greater number of Jupiter Trojans belong to the L4 swarm regardless the value of $A$, which scales the distance from the Lagrangian equilibrium point and thus records information of the  origin and (dynamical and collisional) evolution of Jupiter Trojans.

%The size distribution for L4 and L5 is NOT very similar, because the number ratio at different H is not a constant?? 

Recently, \citet{disi19} studied the long-term stability of the observed Trojans and they found that the escape rate of the L4 population is only 1.1 times smaller than that of the L5 population. In addition, the Yarkovsky effect could also affect the long-term orbital evolution of Jupiter Trojans, but only at the size level of radii $r<1$ km \citep{hell19}. Given a typical albedo of 0.04 \citep{jewi00, fern03, yosh08}, these $r$ values correspond to absolute magnitudes $H>17.6$. As one can see in Fig. \ref{RealJTs}(a), the number asymmetry existing at the brighter region with $H<17$ still remains unexplained. At this point, the number asymmetry between the L4 and L5 Trojan swarms is a real problem, and it should arise at the very beginning of the Solar System when the Jovian planets had not reached their final orbits. 

For the currently observed Jupiter Trojans, the leading-to-trailing number ratio $N_4/N_5$ is $\sim1.8$. However, the value of $N_4/N_5$ is still uncertain since it could be affected by observational biases. With the Trojans detected by the Wide-field Infrared Survey Explorer, \citet{grav11, grav12} estimated a theoretical number ratio of $N_4/N_5=$1.34-1.4. \citet{szab07} made a very careful analysis of the samples listed in the 3rd release of the Sloan Digital Sky Survey, by involving selection effects, sample size and photometry ability. Then, they confirmed that there are significantly more Trojans in the L4 swarm than in the L5 one and obtained a ratio of $N_4/N_5=1.6\pm0.1$. Accordingly, in  this paper we will consider $N_4/N_5=1.6$ as the unbiased value, which is used to characterise the extent of the number asymmetry of Jupiter Trojans.

Up to now we know only a couple of previous works that tried to explain the L4/L5 number asymmetry based on dynamical reasons. \citet{nesy13} proposed that an extra ice giant suffered a scattering encounter with Jupiter during an instability of the outer Solar System. Subsequently, this ice giant came close to the L5 region and depleted the local Trojans, while the L4 swarm was not affected. If so, the resulting number ratio $N_4/N_5$ could reach a value of about 1.3. However, this asymmetry seems a bit weaker than the observation. In another scenario, \citet{pira19} reproduced the L4/L5 asymmetry via the capture of Trojans in early time. Due to the interaction with the gaseous disk, the proto-Jupiter undergoes migration and growth at the same time and starts to trap more Trojans. In the case of a fast inward migration of Jupiter, the relative drift between Jupiter and the co-orbital particles would induce the excess of particles on the L4 side of the horseshoe orbits. After migration stops, the in situ growth of Jupiter shrinks the horseshoe orbits into tadpole orbits. As a result, the Jupiter Trojans are found to be more populous in the L4 swarm than in the L5 swarm. It is worth noting that, to achieve the observed L4/L5 asymmetry, Jupiter has to migrate inwards more than 3.5 AU. Recent work by \citet{deie22}, however, shows that such large scale migration of Jupiter may cause some problems to the inner Solar System, e.g. the amount of mass implanted into the main belt could be very large, and inconsistent with the current low mass of the main belt.  

In this paper, we will offer a new explanation for the L4/L5 number asymmetry of Jupiter Trojans. Our work is organised as follows: in Sect. 2,  we briefly review the fast migration of Jupiter in different scenarios and propose a new picture for the origin and evolution of Jupiter Trojans. In Sect. 3, we investigate the effect of a Jupiter jumping outwards on the stability of the Trojans and present the results that account for the unbiased number ratio of $N_4/N_5\sim1.6$. Also, the effects of different parameters related with Jupiter and its Trojans on the L4/L5 asymmetry are explored. Finally, in Sect. 4, we summarise the results associated with the jumping-Jupiter model and discuss the caveats in this paper; some potential future work to be carried out is also considered therein. 

%In Sect. 4, also for an outward migrating Jupiter but at much higher speeds due to the close encounter with a free floating planet, we provide another mechanism to explain the current L4/L5 asymmetry. Finally, in Sect. 5, we summarise the results associated with the jumping-Jupiter models, and discuss the third potential mechanism of producing the number asymmetry of Trojans in the case that a free floating planet once invaded the L5 region and dispersed a fraction of the local population.

%--------------------------------------------------------------------
\section{New sketch of the origin and evolution}

The capture of Jupiter Trojans could occur when the Solar nebula was still there at the very early stages of planet formation. As we know, there are four possible mechanisms that have been proposed so far: (\rmnum1) the Trojans were captured in situ by a growing proto-Jupiter. The increase of Jupiter's mass caused expansions of both the L4 and L5 regions, and a large amount of planetesimals nearby were trapped into stable Trojan-like orbits \citep{marz98a, marz98b, flem00}.  (\rmnum2)  The planetesimals could have their orbits decayed sunwards due to the gas drag and approached the L4 and L5 points of proto-Jupiter. As a result, some of them were captured into tadpole orbits \citep{peal93, kary95, kort01}. (\rmnum3) The proto-Jupiter underwent fast inward migration in the gaseous disk (i.e. Type-\Rmnum1 migration) and its 1:1 mean motion resonance captured the Trojan asteroids \citep{pira19}. (\rmnum4) With sudden changes in Jupiter's semi-major axis due to close encounters with another ice giant, the Lagrangian regions radially swept through the leftover planetesimal disk. As a consequence, the Trojans were captured during this early dynamical instability among the giant planets \citep{nesy13}.

%\begin{figure}
% \hspace{0 cm}
%  \includegraphics[width=8.5cm]{Fig3.jpg}
%  \caption{Different amplitudes $\Delta{a}_J$ and timescales $\Delta t$ of migration for jumping-Jupiter scenarios. The blue symbols indicate the results from previous works and the associated limiting speed of $\dot{a}_J=\Delta{a}_J/\Delta t=1.5\times10^{-4}$ AU/yr is plotted by the blue dashed line. For reference, the green dashed line is plotted to show the critical migration speed of Jupiter, $\dot{a}_J^{crit}=3.8\times10^{-3}$ AU/yr, corresponding to the disappearance of librational islands around the L4 point.}
%  \label{jump}
%\end{figure}

Considering though Trojan capture through mechanisms (i), (ii), or (iii) seems to be somehow problematic for a few reasons.  First, from the taxonomic classification point of view, Jupiter Trojans are mainly of the D-type (redder) and P-type (less red)  \citep{grav12}. If they were captured while gas was still present \citep[e.g.][]{flem00,wals11}, they should mostly be of C-type, or maybe even S-type. Secondly, observations have shown that Jupiter Trojans have orbital inclinations up to about $40^{\circ}$-$50^{\circ}$. If there was gas, however, during their capture, the Trojan asteroids would be expected to have much lower inclinations. Although \citet{pira19b} proposed that, during the gas-disk phase, Jupiter grew beyond $\sim$30 AU along with a companion planetary embryo may generate P- and D-type Jupiter Trojans with the correct inclinations, this scenario seems too exotic. Finally, as we mentioned earlier,  a large scale inward migration of Jupiter may result in having too much mass into the main asteroid belt and hence conflicting with present day observations \citep{deie22}.

Thus, mechanism (\rmnum4) proposed by \citet{nesy13} becomes a widely accepted model, because it may naturally resolve the aforementioned problems. Yet, this model also has an issue in resolving the L4/L5 asymmetry as we mentioned earlier. As a matter of fact, after the Trojan precursors had initially formed, they could experience violent evolution as Jupiter was migrating. The planet orbital evolution during the dynamical instability stage of planet-planet scattering could be very chaotic \citep[see][for a review]{nesy18}. When Jupiter had gravitational encounters with the fifth outer planet, which was eventually ejected , its outward migration could achieve a significant increase of the semimajor axis ($a_J$) of the order of tenths of an AU and has a timescale of the order of $10^4$ yrs \citep[see Case 2 in][]{nesy13}. Such an outward jump may potentially cause the known L4/L5 asymmetry of Jupiter Trojans.

\begin{figure}
  \centering
  \begin{minipage}[c]{0.5\textwidth}
  \centering
  \hspace{0cm}
  \includegraphics[width=8.5cm]{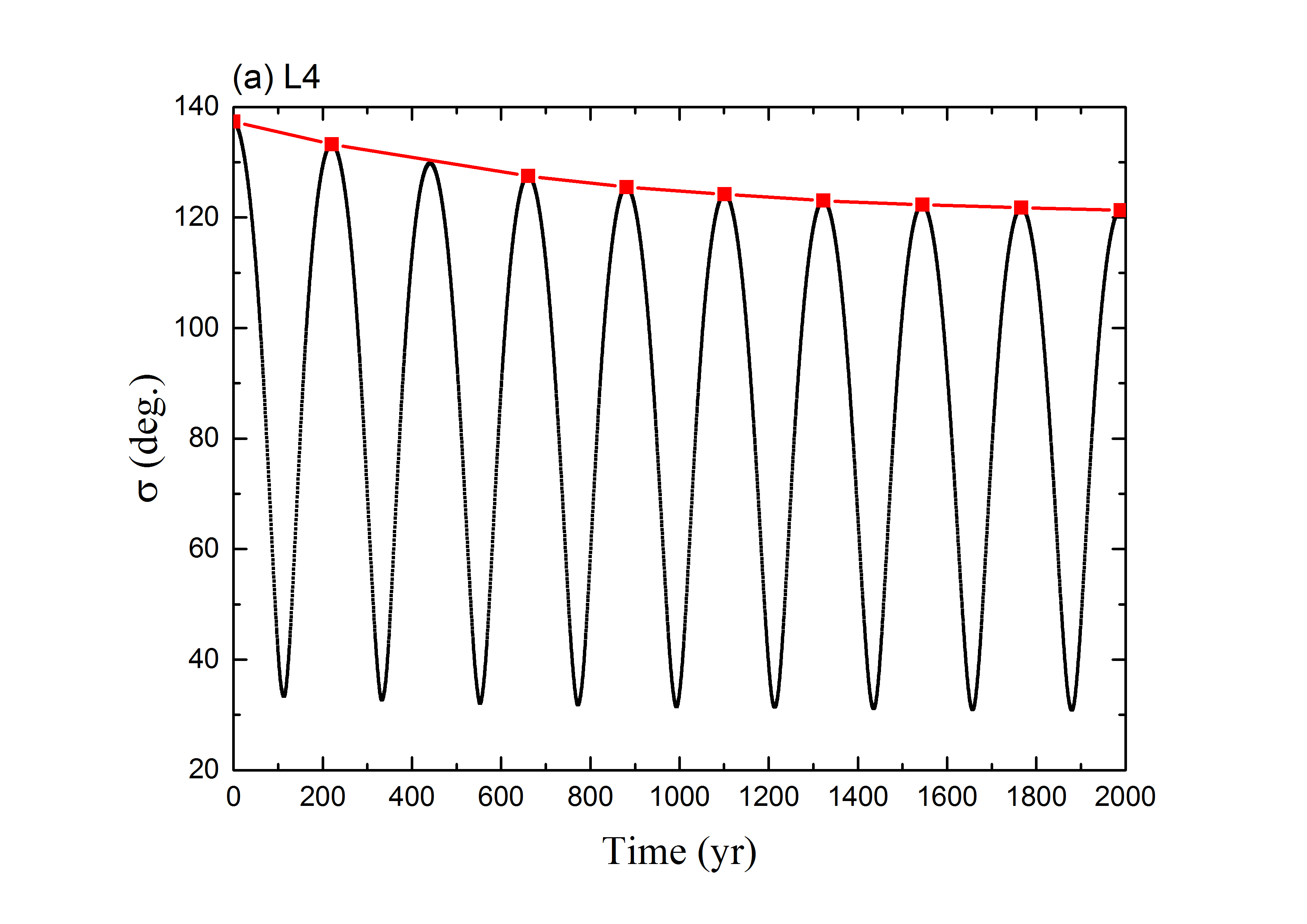}
  \end{minipage}
  \begin{minipage}[c]{0.5\textwidth}
  \centering
  \hspace{0cm}
  \includegraphics[width=8.5cm]{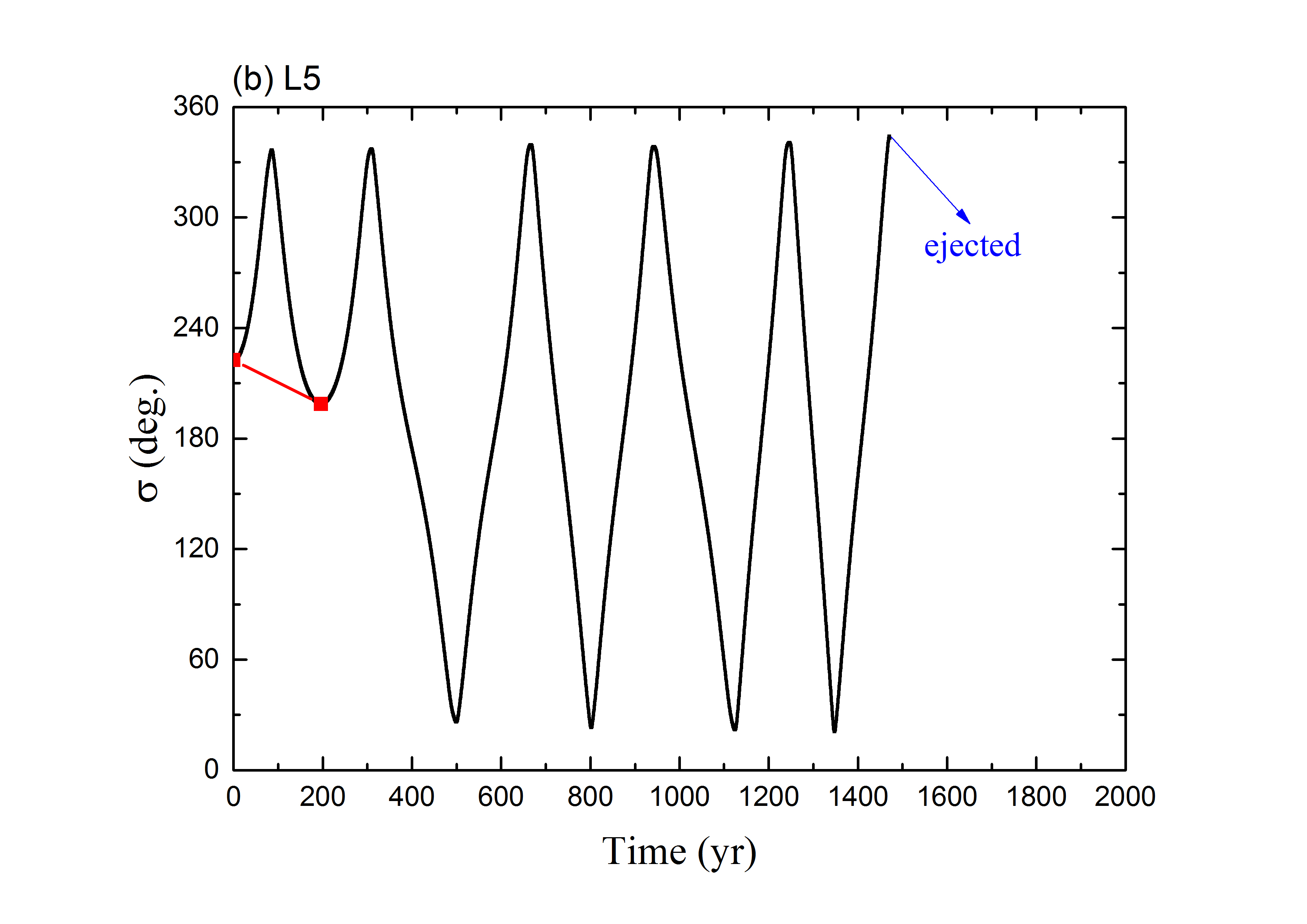}
  \end{minipage}
    \caption{Time evolution of the resonant angle $\sigma$ for the L4 (panel a) and L5 test Trojans (panel b) starting on symmetric orbits, i.e. they initially have the same orbital elements except the mean anomaly, which are determined from $\sigma=137^{\circ}$ and $\sigma=223^{\circ}$ (i.e. $-137^{\circ}$) respectively. In the framework of an outward migration of Jupiter, the L4 sample has its resonant amplitude decreased and becomes more stable, while the L5 one has its resonant amplitude increased and finally gets ejected.}
  \label{migJT}
\end{figure}  

In the framework of the planar circular restricted 3-body problem, \citet{sica03} characterised the distortion of the phase space by a dimensionless parameter $\alpha$. This parameter describes the variation rate of Jupiter's semimajor axis $a_J$ as:
\begin{equation}
\dot{a}_J= (2 a_J n_J \mu ) \cdot \alpha,
\label{speed}
\end{equation}
where $n_J$ is Jupiter's mean motion and $\mu$ is the mass ratio 
\begin{equation}
\mu=\frac{m_J}{m_J+m_{\odot}},
\label{mu}
\end{equation}
with  $m_J$ and $m_{\odot}$ being the masses of Jupiter and the Sun respectively. When Jupiter is still (i.e., $\alpha=0$), the tadpole regions around L4 and L5 are completely symmetric. Considering now an outward migration of Jupiter (i.e. $\alpha>0$), the tadpole region around L4 becomes smaller as $\alpha$ increases, while the tadpole region around L5 is enlarged. For an upper limit value of $\alpha=0.7265$, the L4 and L3 points would merge, and thus the leading librational islands disappear.  By adopting the current values of $\mu=1/1047$, $a_J=5.2$ AU and $n_J=11.8$ yr/2$\pi$, we get a critical migration rate of $\dot{a}_J^{crit}=3.8\times10^{-3}$ AU/yr. The same result was later re-obtained by \citet{ogil06} for the bifurcation of the equilibrium points in co-orbital motion. 
%For reference, this migration limit is plotted as the green curve in Fig. \ref{jump}. In that figure, we can see that the jumping-Jupiter models proposed so far exclusively result in the values of $\dot{a}_J=\Delta{a}_J/\Delta{t}$ below this limit, indicating the persistent existence of the L4 islands and thus the possible survival of the local Trojans. 
In the following, all the considered values of $\dot{a}_J=\Delta{a}_J/\Delta{t}$ are below this limit, indicating the persistent existence of the L4 islands and thus the possible survival of the local Trojans.

Intuitively, since the sizes of the L4 islands would shrink due to Jupiter's outward migration, we expected that the original L4 Trojans with large resonant amplitudes could be left outside the tadpole region and thus become unstable. In order to test this hypothesis, we numerically integrate the orbits of some synthetic L4 Trojans, initially deviating from the L4 point by up to $80^{\circ}$, under the effect of a migrating Jupiter. We adopt the migration amplitude $\Delta a_J=$1 AU and timescale $\Delta t=3.3\times10^3$ yr, i.e. the migration speed of $\dot{a}_J=3\times10^{-4}$ AU/yr. By using Eq. (\ref{speed}) the parameter $\alpha$ is calculated to be about 0.2868, for which the L4 region can still exist \citep{sica03}. In this paper we employ the swift\_rmvs3 symplectic integrator developed by \citet{levi94}, and we use a time-step of 0.5 yr, which is about 1/24 of the orbital period of Jupiter.

To discuss the evolution of Jupiter Trojans, let us introduce the resonant angle
\begin{equation}
  \sigma=\lambda-\lambda_J,
\label{eq:resangle}
\end{equation}
where $\lambda$ and $\lambda_J$ are the mean longitudes of the Trojan and Jupiter respectively. Fig. \ref{migJT}(a) shows a typical example of the evolution of a L4 Trojan starting with the resonant angle $\sigma=137^{\circ}$. It is very interesting to see that, during the outward migration of Jupiter, this object has its maximum resonant angle decreased from $137^{\circ}$ at the beginning to about $120^{\circ}$ at 2000 yr. Hence, the resonant amplitude decreases and the tadpole orbit of this Trojan becomes more stable. This result is contrary to what we conjectured above. We then realise that, in fact, the contraction of the L4 region can push the local Trojans towards the L4 point, i.e. leading them to more stably librating orbits.

Correspondingly, for the L5 Trojans, their resonant amplitudes should increase due to the enlargement of the size of the L5 region. Here, we consider a L5 Trojan starting with the same orbital elements as the discussed L4 example, but its initial resonant angle is set to be $-137^{\circ}$. This way, those two Trojans initially have symmetric resonant configurations as their angular distances from the respective Lagrangian points are both equal to $|\pm 137^{\circ} \mp 60^{\circ}|=77^{\circ}$. As we can see in Fig. \ref{migJT}(b), the L5 example has its minimum resonant angle decreased from $-137^{\circ}$ (i.e. $223^{\circ}$) to about $-159^{\circ}$ (i.e. $201^{\circ}$) within the first libration period, leading to a larger resonant amplitude. In the subsequent evolution, this object crosses the L3 point and switches to a horseshoe orbit centered at $\sigma=180^{\circ}$. At about 1500 yr, the resonant amplitude of the horseshoe orbit approaches a quite large value of $\sim164^{\circ}$ and the considered L5 Trojan eventually suffers Jupiter's strong perturbation that ejects it out of the co-orbital region. 

Based on the above new findings, we suppose that the survival rate of the L4 swarm could be distinctly higher than that of the L5 swarm during the fast outward migration of Jupiter. In this dynamical evolution, the extent of the number asymmetry between these two populations essentially depends on Jupiter's migration speed $\dot{a}_J$. Given an appropriate value of $\dot{a}_J$, this mechanism may account for the unbiased asymmetry of $N_4/N_5=1.6$ that we required before.

The aim of this work is to evaluate the effect of an outward migrating Jupiter on the L4/L5 asymmetry. While we have to mention that, after an outward giant leap, Jupiter could experience successive migrations due to close encounters with large planetesimals. Such evolution would only induce small variations of Jupiter's semimajor axis, somewhat analogous to the grainy migration of Neptune \citep{nesv16}, only by an extent of $\Delta a_J\lesssim0.1$ AU. Since the associated migration timescale is generally at least of the order of $10^4$ yr, as we find in the following simulations, such a slow migration of Jupiter would contribute little to the current asymmetry problem of Jupiter Trojans. 

%We would like to mention that, probably Jupiter once experienced an outward large jump due to a very close encounter with an additional ice giant \citep{nesy13}. After that, successive medium or distant encounters between Jupiter and this ice giant could occur, as well as close encounters between Jupiter and large planetesimals in the disk. These processes would induce small variations of Jupiter's semimajor axis $a_J$, somewhat analogous to the grainy migration of Neptune \citep{nesv16}. Since the perturbation time is generally at least on the order of 1000 yr, as we show later, a value of $a_J<0.1$ AU would contribute little to the asymmetry problem of Jupiter Trojans. 

%______________________________________________________________

\section{Number asymmetry of L4 and L5 Trojans}

In this section, we consider a simplified Solar System that consists of the Sun, Jupiter and a number of ``captured'' Trojans around the L4 and L5 points. We assume that after the capture, the two Trojan swarms have the same number of objects and their orbital distributions are nearly symmetric. In the subsequent numerical orbital calculations with or without the migration of Jupiter, we employ again the swift\_rmvs3 integrator with a time-step of 0.5 yr.

\subsection{Pre-runs and test Trojans}

In order to mimic the ``captured'' Trojans, denoted by test Trojans hereafter, we conduct a series of pre-runs lasting 2000 yr, within which Jupiter is fixed on its currently observed orbit with $a_J=5.2$ AU. Since the Jupiter Trojans librate around the L4 or L5 points with periods of a few hundred years, the 2000 yr evolution is long enough to identify the tadpole motion. Initially, all particles have the same semimajor axis ($a=5.2$ AU) with Jupiter, and their eccentricities $e$ are randomly selected within the interval ranging from 0-0.3. Here, the initial inclinations of the Trojans are taken to be 0.01, which is consistent with the theoretical coplanar 3-body model constructed in Sect. 2. Actually, the scenario by \citet{nesy13} could produce high inclination Trojans, which will be considered later. As for the other three orbital elements of each particle, the longitude of ascending node $\Omega$ and argument of perihelion $\omega$ are chosen randomly between $0^{\circ}$ and $360^{\circ}$, while the mean anomaly $M$ is determined from the initial resonant angle $\sigma_0$. 

For the L4 and L5 swarms, the values of $\sigma_0$ are given by $60^{\circ}+\Delta \sigma_0$ and $-60^{\circ}-\Delta \sigma_0$ respectively, where $\Delta \sigma_0$ refers to a certain range. 
%For the known Trojans, our previous work shows the range $\Delta \sigma_0=0^{\circ}$-$50^{\circ}$ \citep{li2018}. 
The parameter $\Delta \sigma_0$ characterises the displacements of Trojans from the individual Lagrangian points and different ranges are to be considered later. The choice of $\Delta \sigma_0$ would influence the survivability of the Trojans even in the short-term pre-runs and we would accordingly adjust the total number ($N_t$) of particles to generate a fixed number ($N_c$) of test Trojans to be used for the jumping-Jupiter model. Generally, we have to choose $N_t$ at least three times larger than $N_c$ in order to obtain a near uniform distribution of initial resonant angles of test Trojans.

%For the L4 and L5 swarms, the values of $\sigma$ are randomly chosen in $[60^{\circ}+\Delta \sigma_1, 60^{\circ}+\Delta \sigma_2]$ and $[-60^{\circ}-\Delta \sigma_1, -60^{\circ}-\Delta \sigma_2]$, respectively.

After some experiments, we restrict ourselves to consider only the rather stably librating particles with $\Delta\sigma_0\le80^{\circ}$ in the pre-runs. Given a specific $\Delta\sigma_0$, the number $N_c$ of test Trojans is set to be 500 for each swarm separately. In the following study of the L4/L5 asymmetry, this $N_c$ is neither too small for the statistics of survivals, nor too large for a reasonable computational time. 

\subsection{Jumping-Jupiter scenario}
 
As we described in Sect. 2.1, it is possible that Jupiter made an outward jump in the early stages of the Solar System due to a close encounter with the fifth outer planet. According to \citet{nesy13}, the migration amplitude $\Delta a_J$ could reach values as large as a few tenths of an AU on a timescale $\Delta t$ of the order of $10^4$ yr. Comparable values of $\Delta a_J$ and $\Delta t$ will be explored within the range of $\dot{a}_J=\Delta{a}_J/\Delta{t}<1.5\times10^{-4}$ AU/yr, where the L4 islands can persistently exist.

% As we described in Sect. 2.1, it is possible that Jupiter made an outward jump in the early stages of the Solar System, due to either the interaction with the gaseous disk or a close encounter with an ice giant. The related results in previous works about Jupiter's fast migration are listed in Fig. \ref{jump}, as indicated by the blue symbols. The migration amplitudes $\Delta a_J$ could reach values as large as 0.5-2 AU, on timescales $\Delta t$ of several thousand years. Accordingly, the dashed blue curve is plotted in Fig. \ref{jump} to represent a limiting speed of about $1.5\times10^{-4}$ AU/yr, which indicates what ranges of $\Delta a_J$ and $\Delta t$ could be explored.

%Fig. \ref{jump} shows what range of Jupiter's migration distance () and timescale could be explored, the blue symbols represent   

Hereafter, we investigate the effect of such a rapid outward migration of Jupiter on the difference of the surviving rates between the L4 and L5 Trojan swarms. Although the migration scenario of Jupiter is quite delicate, in general it could be simply described by a time variation of Jupiter's semimajor axis \citep{malh95}:
\begin{equation}
  a_J(t)=a_J+\Delta a_J [1-\exp(-t/\tau)],
\label{eq:variation}
\end{equation}
where $a_J$ is Jupiter's current semimajor axis of 5.2 AU, $a_J(t)$ is the value at epoch $t$ and $\Delta a_J$ is the migration amplitude as we noted above. The e-folding time $\tau$ can quantitatively measure the total migration time $\Delta t$. We here define a correlation by 
\begin{equation}
\Delta t = (10/3)~\tau.
\label{taoTOdt}
\end{equation}
Because at the epoch of $t=(10/3) \tau$, Jupiter's asymptotic migration rapidly slows down; and, Jupiter has travelled over a distance of $0.96\Delta a_J$, which is approximate to the total migration amplitude of $\Delta a_J$.

To implement the variation of $a_J(t)$ described by Eq. (\ref{eq:variation}) in the swift\_rmvs3 integrator, we add an artificial force on Jupiter along the direction of its orbital velocity $\mathbf{\hat{\nu}}$, as
\begin{equation}
  \Delta \mathbf{\ddot{r}} =\frac{\mathbf{\hat{\nu}}}{\tau}
  \left\{\sqrt{\frac{GM_\odot}{a_J}}-\sqrt{\frac{GM_\odot}{a_J+\Delta a_J}} \right\}
  \exp\left(-\frac{t}{\tau}\right),
\end{equation}
where $M_\odot$ is the mass of the Sun, and $G$ is the gravitational constant.

\subsection{General results}

\begin{table}
 \hspace{-1 cm}
\centering
\begin{minipage}{8cm}
\caption{The statistics of survived test Trojans for different initial resonant angles $\sigma_0$ within the \textit{control} model of Jupiter migration (i.e. amplitude $\Delta a_J=0.5$ AU and  e-folding time $\tau=1000$ yr). For each $\Delta\sigma_0$, there are 500 L4 samples with $\sigma_0$ in the range of $60^{\circ}+\Delta \sigma_0$, and 500 L5 samples with symmetric $\sigma_0$ in the range of $-60^{\circ}-\Delta \sigma_0$. After the 1 Myr evolution, we derive the numbers ($N_4$, $N_5$) and the final resonant amplitudes ($A_4$, $A_5$) of survived Trojans, with subscripts `4' and `5' for the L4 and L5 swarms respectively. The number ratio $R_{45}=N_4/N_5$ is presented in the last column.}      % title of Table
\label{model1}
\begin{tabular}{c| c c c c c}        % centered columns (9 columns)
\hline                 % inserts double horizontal lines
$\Delta\sigma_0$    &         $N_4$      &        $A_4$          &         $N_5$           &      $A_5$       & $R_{45}$                \\

\hline

 0-80                            &          399            &        1-51            &         350                &        1-51               &    1.14           \\
           
10-80                            &          394            &        2-51            &         309                &        3-50               &    1.27           \\

20-80                            &          356            &        8-52            &   277                       &        11-52       &        1.29           \\

\textbf{30-80 }          &          \textbf{312}            &      \textbf{15-54}             &         \textbf{209}                &        \textbf{18-56}       &        \textbf{1.50}           \\

30-70                           &          356            &       10-46            &         256                 &        10-52       &       1.39           \\

\hline
\end{tabular}
\end{minipage}
\end{table}

In this subsection, in the framework of jumping Jupiter, a series of runs will be performed by varying three key parameters:\\ 
(1) the amplitude $\Delta a_J$ of Jupiter's migration,\\
(2) the e-folding time $\tau$ for Jupiter's migration. For the sake of convenience, we use $\tau$ in the numerical integration, while the migration timescale can be obtained by multiplying $\tau$ by 10/3 (see Eq. (\ref{taoTOdt})).\\
(3) The initial displacements $\Delta \sigma_0$ of the test Trojans from the associated Lagrangian point.

First, we consider the jumping-Jupiter model with $\Delta a_J=0.5$ AU and $\tau=1000$ yr (i.e. $\Delta t\sim 1/3\times 10^4$ yr). This will be denoted as the \textit{control} migration model. We now focus on the dependence of the L4/L5 asymmetry on the distribution of the initial resonant angles $\sigma_0$ of the test Trojans. The results are presented in Table \ref{model1}. The first column lists the selected $\Delta \sigma_0$ within the stability limit of $\le80^{\circ}$, with the minimum value increasing as we move to the bottom of the table. Taking $\Delta \sigma_0=10^{\circ}$-$80^{\circ}$ as an example, generated from the pre-runs, there are 500 L4 test Trojans with initial resonant angles $\sigma_0$ distributed nearly uniformly between 70$^{\circ}$ and 140$^{\circ}$, and 500 L5 test Trojans with $\sigma_0=220^{\circ}$-$290^{\circ}$. We then integrate the system for 1 Myr which is long enough to allow Jupiter Trojans to experience the secular orbital precession after Jupiter's migration is over. At the end of the integration, the rather stable Trojans can be obtained. 

In Table \ref{model1}, the second and third columns indicate, for the survived L4 population, the number ($N_4$) and the final resonant amplitude ($A_4$) distribution respectively. The number ($N_5$) and resonant amplitudes ($A_5$) of the survived L5 Trojans are given in the next two columns. In the last column, the key result is depicted, that the obtained number ratios $R_{45}$($=N_4$/$N_5$) are clearly  larger than 1. For the first four cases with $\Delta\sigma_0$ up to $80^{\circ}$, the ratio $R_{45}$ increases from 1.14 for minimum $\Delta\sigma_0=0$ to 1.50 for minimum $\Delta\sigma_0=30^{\circ}$. This trend is straightforward for the reason that the test Trojans with large $\sigma_0$ would be higher in proportion and the number asymmetry is mainly caused by this population. Similarly, for the last two cases, by choosing the same lower limit of $\Delta\sigma_0=30^{\circ}$ but reducing the maximum $\Delta\sigma_0$ from $80^{\circ}$ to $70^{\circ}$, the ratio $R_{45}$ falls from 1.50 to 1.39 as a result of fewer particles with largest $\sigma_0$. As a matter of fact, the choice of $\Delta\sigma_0$ could be much different, not only in the range, but also in the profile of the distribution. Here, we just want to put forward a possible explanation for the unbiased ratio of the number of the L4 swarm to the number of the L5 swarm, at the level of 1.6. In what follows, we will consider $\Delta\sigma_0=30^{\circ}$-$80^{\circ}$ to be the \textit{standard} setting, and the ratio $R_{45}=1.50$ obtained here (second last row in Table \ref{model1}, in bold) would continue to increase in the longer Gyr evolution.

%As for each set of 500 test Trojans, they are sampled on the tadpole orbits with a particular $\Delta \sigma_0$ listed the first column of Table \ref{model1}. For example, given $\Delta \sigma_0=$10-$80^{\circ}$, the sample Trojans have initial resonant angles $\sigma$ distributed nearly uniformly between 70$^{\circ}$ and 140$^{\circ}$.

\begin{figure}
  \centering
  \begin{minipage}[c]{0.5\textwidth}
  \centering
  \hspace{0cm}
  \includegraphics[width=8.5cm]{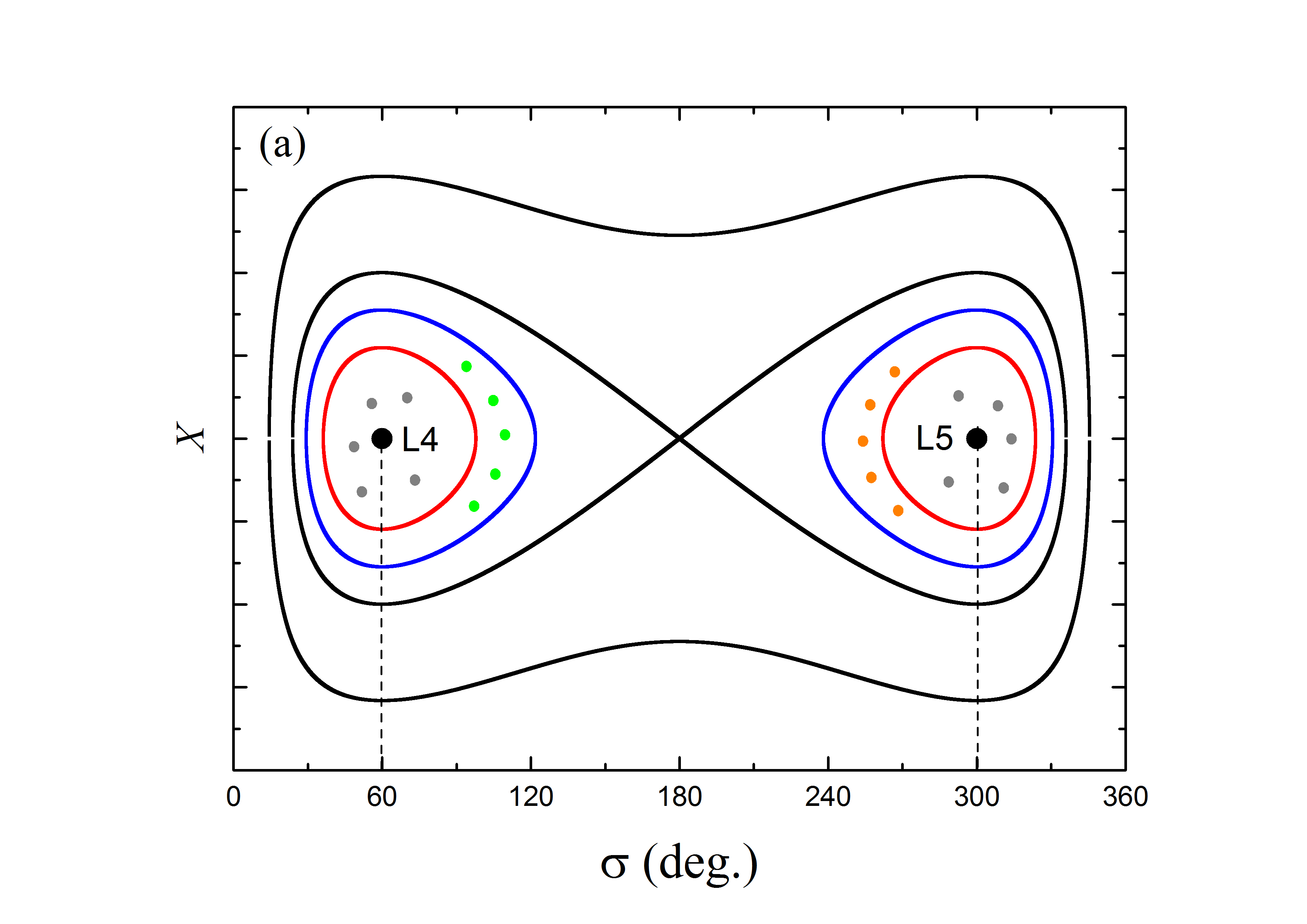}
  \end{minipage}
  \begin{minipage}[c]{0.5\textwidth}
  \centering
  \hspace{0cm}
  \includegraphics[width=8.5cm]{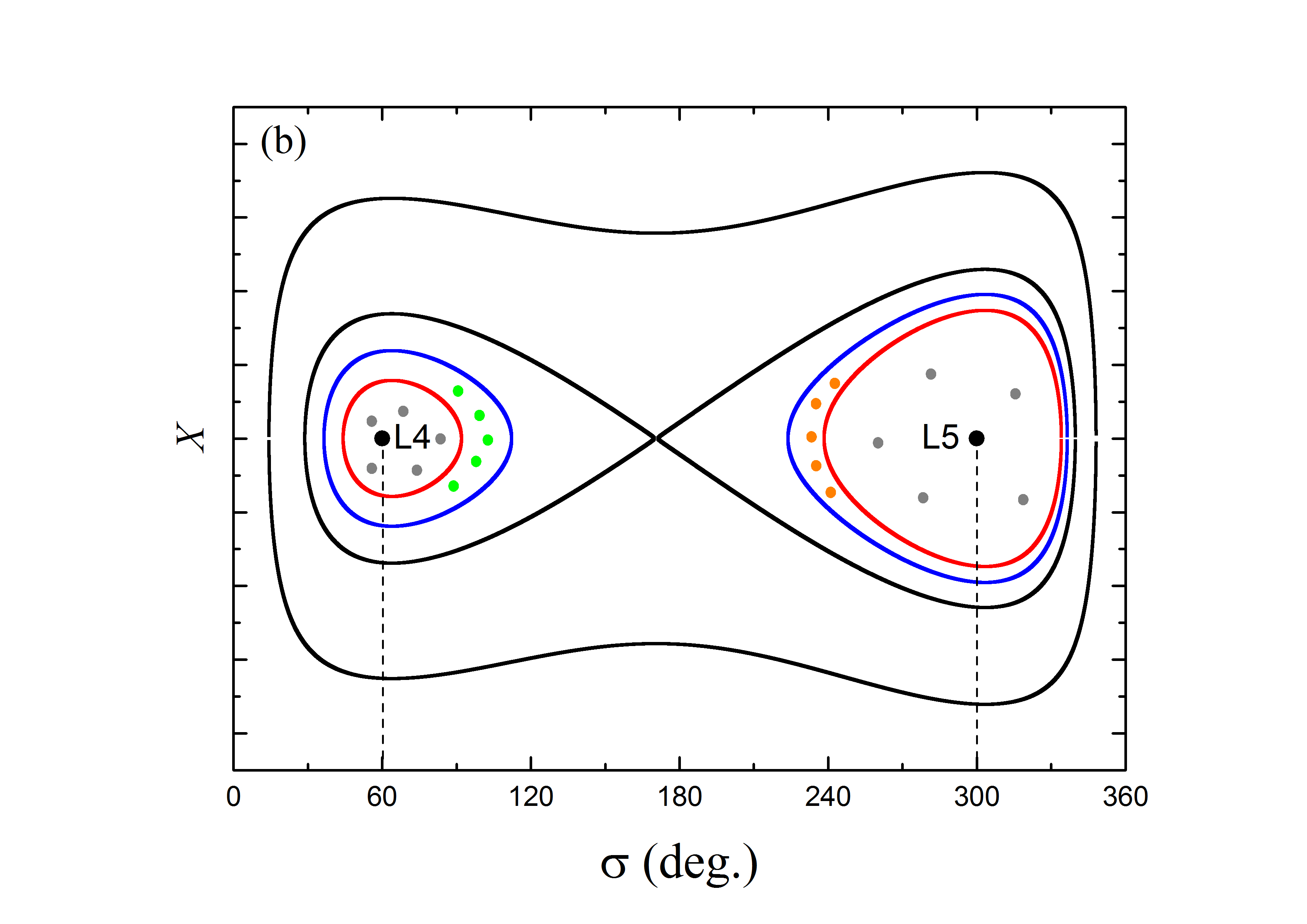}
  \end{minipage}
    \caption{The phase space of the co-orbital motion near Jupiter's L4 and L5 points for the non-migrating Jupiter (panel (a)) and an outward migrating Jupiter with a speed of $1.5\times10^{-4}$ AU/yr (panel (b)). This figure is generated by the theory developed in \citet{sica03}, and the ordinate denotes the parameter $X$ that is proportional to the distance of the Trojan with respect to Jupiter. The blue curve represents the stability limit of Jupiter Trojans, and the adjacent green (around L4) and orange (around L5) dots indicate the Trojans with large resonant amplitudes. Interior to the red curve, the Trojans (grey dots) with smaller resonant amplitudes are more stable. During the outward migration of Jupiter, the L4 Trojans drift toward the L4 point due to the contraction of the librational islands, while the L5 Trojans move away from the L5 point due to the expansion of the librational islands. For reference, the vertical dashed lines are plotted corresponding to the L4 and L5 points respectively.}
  \label{phase}
\end{figure}

\begin{figure}
  \centering
  \begin{minipage}[c]{0.5\textwidth}
  \centering
  \hspace{0cm}
  \includegraphics[width=8.5cm]{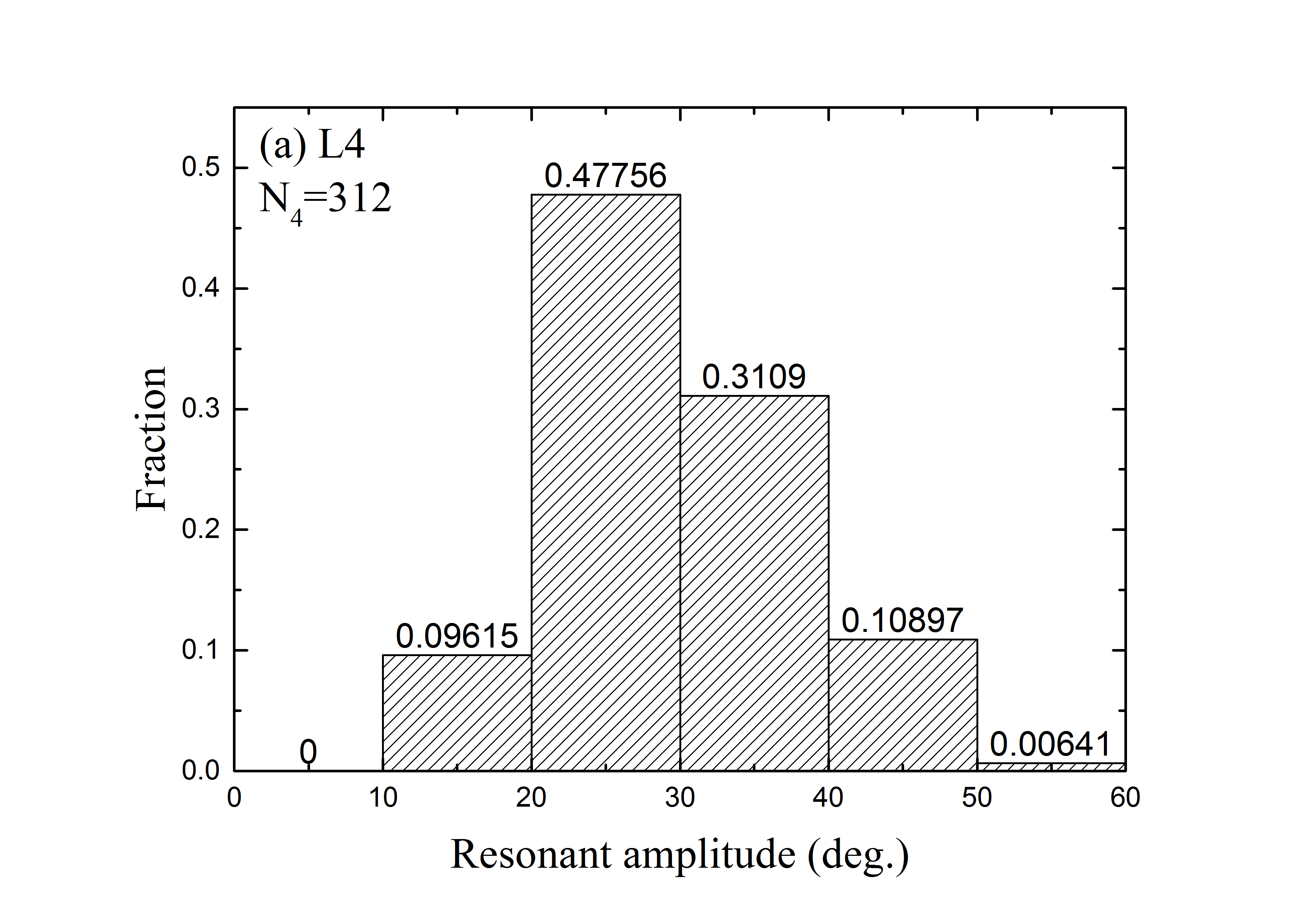}
  \end{minipage}
  \begin{minipage}[c]{0.5\textwidth}
  \centering
  \hspace{0cm}
  \includegraphics[width=8.5cm]{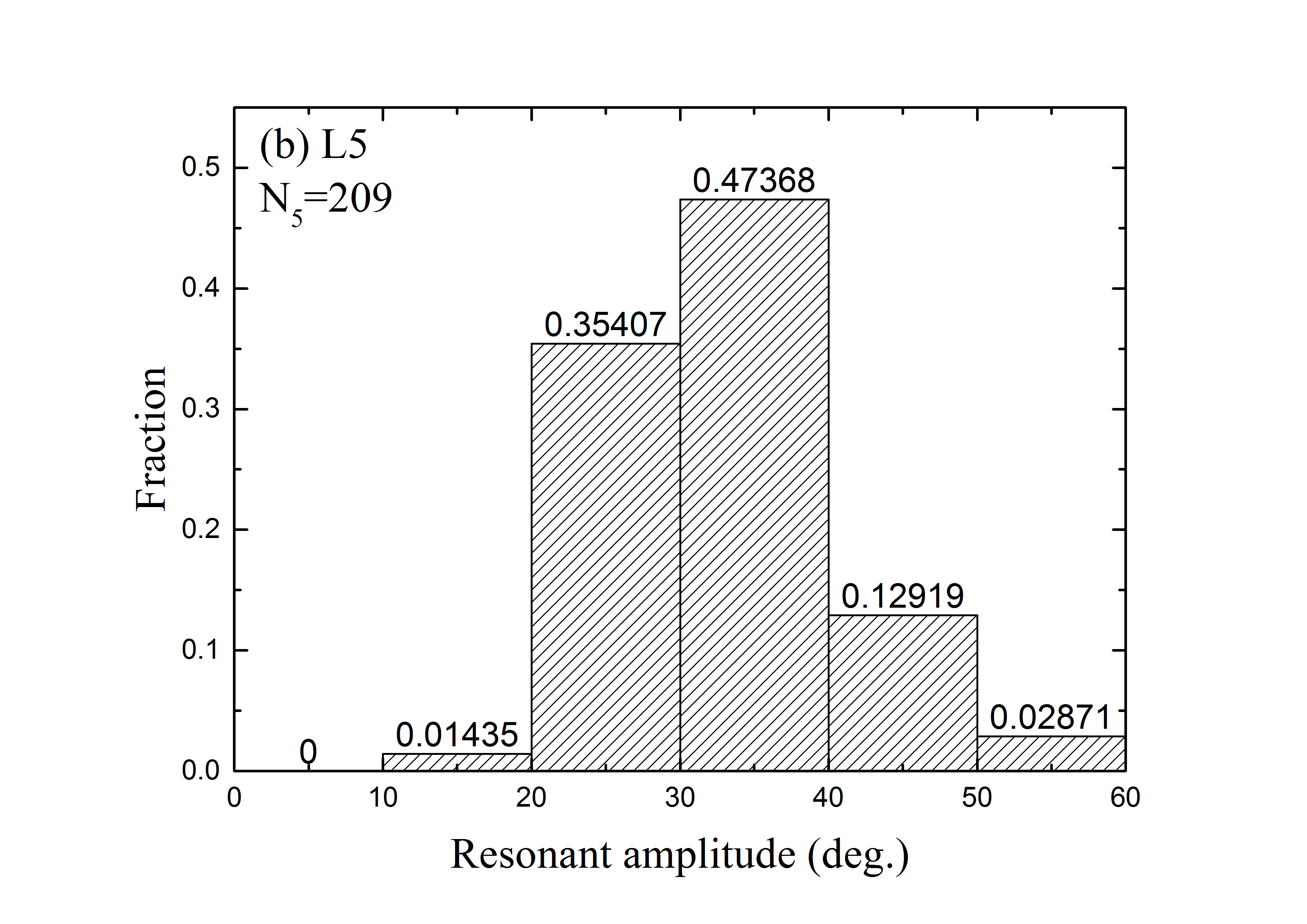}
  \end{minipage}
    \caption{Distribution of resonant amplitudes of survived test Trojans, for the case of the standard setting of $\Delta\sigma_0=30^{\circ}$-$80^{\circ}$ in the control model. The corresponding statistical results can be found in the row in bold in Table \ref{model1}.}
  \label{Adistribution}
\end{figure}

A schematic illustration of the different evolution paths of the L4 and L5 Trojans caused by the jumping Jupiter is provided in Fig. \ref{phase}. Using the theory developed by \citet{sica03}, we construct the phase space of the co-orbital motion near Jupiter's L4 and L5 points for several values of the Jacobi constant. For the non-migrating case shown in panel (a),  we denote the blue curve to be the stability limit of Jupiter Trojans, and the red curve is drawn nearby. An outward migration of Jupiter could induce the contraction of the L4 region and the expansion of the L5 region, as the variation of the red (or blue) curve in panel (b) versus that in panel (a). As a result, for the Trojan populations with large resonant amplitudes, the L4 swarm (green dots) would be pushed towards the L4 point and their resonant amplitudes $A_4$ decrease, while the L5 swarm (orange dots) would be driven even far away from the L5 point and their resonant amplitudes $A_5$ increase, similarly to the two examples shown in Fig. \ref{migJT}. After the migration has completely ceased, the L5 region is restored to its original size, i.e. becoming smaller, as the colourful curves around L5 in panel (b) change back to those in panel (a). That means that those L5 Trojans with excited $A_5$, as indicated by the orange dots in panel (b), could be outside the stability limit denoted by the blue curve in panel (a). Thus, they would continue to be ejectable, leading to even fewer survivals around the L5 point. But due to the decrease in $A_4$, the L4 Trojans indicated by the green dots have reached more stable orbits, leading to more L4 survivals. Therefore, we finally get a number ratio $R_{45}$ larger than 1.

Apart from the Trojans with large resonant amplitudes between the red and blue curves in Fig. \ref{phase}, we now turn to the ones (grey dots) inside the red curves. These objects have medium to small resonant amplitudes, and they should also be affected by the distortion of the red curves in the phase space. Similarly, the grey samples around the L4 point would have their resonant amplitudes $A_4$ decreased, and the grey ones around the L5 point would have their resonant amplitudes $A_5$  increased. Although the L4 and L5 grey swarms can both survive to the end of the integration, the former may have relatively smaller resonant amplitudes. This could be validated by the results in Table \ref{model1}, that the minimum $A_4$ is generally smaller than the minimum $A_5$. However, as the difference seems not to be very significant, we suppose that it may be due to the slow diffusion of the Trojans after Jupiter's migration era, in the numerical integration of 1 Myr in total. Probably, we should take an overall look at the resonant amplitude distribution.

%Taken in total, the survived L4 swarm should have relatively smaller resonant amplitudes than the L5 swarm. However, we notice in the Table \ref{model1} that, the ranges of $A_4$ and $A_5$ determined from the last $10^5$ yr evolution are generally comparable. The seeming conflict is because the occasional diffusion after Jupiter migration era, in the numerical integration of 1 Myr in total.

For the standard setting of $\Delta\sigma_0$ (i.e. $30^{\circ}$-$80^{\circ}$), Table \ref{model1} gives comparable ranges of $A_4=15^{\circ}$-$54^{\circ}$ and $A_5=18^{\circ}$-$56^{\circ}$ for the survived L4 and L5 swarms respectively. We then plot the distributions of the resonant amplitudes for these two populations, as shown in Fig. \ref{Adistribution}. It is quite obvious that, in either of the bins with smaller resonant amplitudes of $10^{\circ}$-$20^{\circ}$ and $20^{\circ}$-$30^{\circ}$, the L4 population has a larger fraction of survived Trojans than the L5 population. Accordingly, for objects having resonant amplitudes smaller than $30^{\circ}$, the fractions of survived Trojans of the L4 and L5 swarms are $\sim57$ per cent and $\sim37$ per cent respectively. Therefore, from such a statistical point of view, the numerical result is in good agreement with what we theoretically expected above. 

\begin{table}
 \hspace{-1 cm}
\centering
\begin{minipage}{8cm}
\caption{Similarly to Table \ref{model1}, but only for the standard setting of $\Delta\sigma_0=30^{\circ}$-$80^{\circ}$. We extend the runs to a total of 1 Gyr of integration time. The first row corresponds to the same case highlighted in bold in Table \ref{model1}.}      % title of Table
\label{standard}
\begin{tabular}{c| c c c c c}        % centered columns (9 columns)
\hline                 % inserts double horizontal lines
Time (Myr)       &         $N_4$      &        $A_4$          &         $N_5$           &      $A_5$      & $R_{45}$                  \\

\hline
           
\textbf{1}            &          \textbf{312}                &      \textbf{15-54}             &         \textbf{209}              &        \textbf{18-56}       &        \textbf{1.50}           \\

10                          &          284           &   15-50      &        185       &   18-52       &         1.54                      \\

100                        &          268            &   12-49     &        161       &   18-52       &         1.66                       \\

1000                     &          255             &   14-49    &        151       &   16-50       &         1.68                       \\

\hline
\end{tabular}
\end{minipage}
\end{table}

\begin{table*}
 \hspace{-2 cm}
\centering
\begin{minipage}{13 cm}
\caption{Similarly to Table \ref{model1}, but only for the standard setting of $\Delta\sigma_0=30^{\circ}$-$80^{\circ}$. We adopt different migration rates of Jupiter, $\dot{a}_J=\Delta a_J / [(10/3) \tau]$. In this table, we only list the L4/L5 number ratios $R_{45}$ for survived Trojans (fractions in parentheses are $N_4/N_5$). The bold $\Delta a_J$ and $\tau$ are the same parameters that we used in Table \ref{model1} and consequently the bold $R_{45}$ value indicate the same results, i.e. for the control model.}      % title of Table
\label{aJTao}
\begin{tabular}{r| l c c c l}        % centered columns (9 columns)
\hline                 % inserts double horizontal lines

%\diagbox{$\tau$}{$\Delta a_J$}    &     0.5 AU                                             &      0.1 AU          &     0.25 AU         &     1 AU               &     2 AU                    \\

%\hline
           
%$10^3 yr$                                               &          \textbf{1.50(312/209)}           &     1.06               &        1.29             &         2.08             &          3.18(496/156)     \\    

%$0.5\times10^3 yr$                             &          2.36                                             &     1.22               &        1.52             &         3.54             &          3.72                       \\      

%$2\times10^3 yr$                                &          1.15                                             &     1.01               &         1.17             &        1.44              &          1.88                \\     

%$4\times10^3 yr$                                &          1.11(283/255)                           &     1.03               &         1.04             &        1.17                  &         1.43                \\     

\diagbox{$\tau$}{$\Delta a_J$}    &     0.25 AU              &        \textbf{0.5 AU}                   &     1 AU               &     2 AU                    \\

\hline

$0.5\times10^3$ yr                             &        1.52                  &             2.36                                    &         3.54             &          3.72                       \\  
           
\textbf{10$^3$ yr }                             &         1.29                 &       \textbf{1.50(312/209)}        &         2.08             &          3.18(496/156)     \\    

$2\times10^3$ yr                                &         1.17                 &          1.15                                        &        1.44             &          1.88                \\     

$4\times10^3$ yr                                &         1.04                  &          1.11(283/255)                     &        1.17             &         1.43                \\   

\hline
\end{tabular}
\end{minipage}
\end{table*}

To further check the temporal variation of the L4/L5 asymmetry on a timescale of the order of the age of the Solar System, for the case with standard $\Delta\sigma_0$ in the control migration model (i.e. the bold row in Table \ref{model1}), we extend the integration up to 1 Gyr. As before, we evolve our system using Newtonian gravity only, while the Yarkovsky effect is not considered because it has little contribution for  objects with radii larger than 1 km. Recorded in Table \ref{standard}, the leading-to-trailing number ratio $R_{45}$ increases with time, from 1.50 at 1 Myr to 1.68 at 1 Gyr. In addition, over a period from 100 Myr to 1 Gyr, we notice that the increase of this ratio is as small as only about 1\%. Thus we believe that $R_{45}$ has nearly converged to the limiting value and it will not change significantly if the integration was carried on for an even longer time (e.g. 4.5 Gyr). We additionally note that, starting from 1 Myr, there are 255 out of 312 L4 Trojans (i.e. 81.7\%) that could survive up to 1 Gyr, while 151 out of 209 L5 Trojans (i.e. 72.2\%) could have survived. This implies that the surviving rate of the L4 swarm is about 1.13 times greater than that of the L5 swarm. This number qualitatively agrees with the result of \citet{disi14}, indicating that the giga-year evolution may contribute an additional $\sim10$\% asymmetry after Jupiter Trojans had evolved to their current locations. For the sake of saving computational time, in what follows we will only carry out the integration for 1 Myr in order to obtain the number ratio $R_{45}$ in rough, and the final value of $R_{45}$ could be simply increased by a factor of 1.1. Accordingly, to achieve the asymmetry of $R_{45}\sim1.6$, we would require a value at the level of $>1.4-1.5$ (i.e. $\gtrsim 1.6/1.1$) in the 1 Myr simulation.

As we anticipated above, the faster (slower) Jupiter's outward migration is, more (fewer) L4 Trojans could survive, while fewer (more) L5 ones do, leading to a larger (smaller) value of the number ratio $R_{45}$. Surely, we need to keep the migration speed under the limit of $\dot{a}_J^{crit}=3.8\times10^{-3}$ AU/yr, otherwise the islands around L4 would disappear. For the same test Trojans with standard $\Delta\sigma_0$ used in the previous simulations, we would test a variety of distances $\Delta a_J$ and e-folding times $\tau$ of Jupiter's migration within the plausible ranges. In Table \ref{aJTao}, we present the results of a series of Jupiter migration scenarios. The control model is indicated by bold characters and one can see the corresponding results reported in Tables \ref{model1} and \ref{standard} (also highlighted in bold). We find that: (1) The diagonal values of $R_{45}$ correspond to the same migration rate of $\dot{a}_J=\Delta a_J / [(10/3) \tau]=1.5\times10^{-4}$ AU/yr as that adopted in the control migration model. These number ratios fall in the region of 1.43-1.53, in good agreement with the result of the control model (i.e. $R_{45}=1.5$) and could account for the current asymmetry problem. (2) The upper right values of $R_{45}$ correspond to faster migration of Jupiter, i.e. $\dot{a}_J=$(3-12)$\times10^{-4}$ AU/yr. As expected, we can obtain larger $R_{45}$ up to 3.72, and this number ratio could  continue to increase with $\dot{a}_J$. However, whether a too fast migration of Jupiter is reasonable or not may need further investigation of the model of planet-planet scatterings. Here, we need to note that for achieving a certain $R_{45}$, the larger $\dot{a}_J$ can allow the lower limit of $\Delta\sigma_0$ to be smaller value. Thus, the coupled contribution from the migration of Jupiter and the initial distribution of the Trojans may place further constraints on the capture process of Jupiter Trojans occurred at an earlier time. (3) On the other hand, the lower left part indicates a slower migration of Jupiter and the required condition of $R_{45}>$1.4-1.5 can not be satisfied. (4) In the case of $\tau=10^3$ yr, when we increase $\Delta a_J$ from 0.5 AU used in the control model to 2 AU (i.e. Jupiter's migration speed becomes 4 times larger), 496 out of 500 L4 test Trojans could survive. This suggests that the L4 region becomes very stable that nearly all the local Trojans, even those starting with quite large resonant amplitudes, could survive. Thus we have that the number $N_4$ of L4 Trojan is about a constant. While for the L5 Trojans, the number of survivals would decrease monotonically with larger $\Delta a_J$, i.e. $N_5$ keeps decreasing. As a result, the number ratio $R_{45}=N_4/N_5$ could be significantly larger than 1. This additionally supports the effect of a rather fast migration of Jupiter. 

It is worth mentioning that, in the framework of the jumping-Jupiter model, all the simulated Trojans would end with resonant amplitudes $\gtrsim15^{\circ}$, which are in general higher than those of the observed Trojans (see Fig. \ref{RealJTs}(b)). A possible mechanism is that collisions among Trojans may reduce their resonant amplitudes \citep{marz98b, marz00}. The collisional evolution of Jupiter Trojans after the migration of Jupiter still needs to be reassessed in the future. This may allow us to constrain the orbital distribution of the primordial Trojans and the process of the subsequent Jupiter's jump.

%\textbf{Theoretically speaking, the resulting L4/L5 asymmetry could be even more significant as long as the migration of Jupiter is faster, as we can see in the last four cases shown in Table \ref{aJTao}. It is worth mentioning that, the larger the migration speed $\dot{a}_J$, the larger the final resonant amplitudes $A_4$ and $A_5$. Indeed, in the considered jumping Jupiter model, all the simulated Trojans would end with resonant amplitudes $\gtrsim15^{\circ}$, which are in general higher than those of the observed Trojans (see Fig. \ref{RealJTs}(b)). A possible mechanism is that collisions among Trojans may reduce their resonant amplitudes \citep{marz98b, marz00}. The collisional evolution of Jupiter Trojans after the migration of Jupiter still needs to be reassessed in the future. This may allow us to constrain the orbital distribution of the primordial Trojans and the process of the subsequent Jupiter's jump.}

In summary, given a proper distribution of primordial Jupiter Trojans after the capture, the fast outward migration of Jupiter is capable of inducing a leading-to-trailing number ratio $R_{45}>1$. Consequently, this dynamical mechanism can well explain the unbiased asymmetry of $R_{45}\sim1.6$ based on the current observation. Theoretically speaking, the resulting L4/L5 asymmetry could be even more significant as long as the migration of Jupiter is fast enough. More interestingly, if the jumping-Jupiter has a speed $\dot{a}_J$ higher than the critical value $\dot{a}_J^{crit}$, the L4 point will disappear.

\subsection{Effects of the system parameters}

It is likely that the system parameters in the jumping-Jupiter model could relate to the L4/L5 asymmetry. We should figure out how different parameters associated with Jupiter and its Trojans would affect our results obtained above. In what follows, we will explore the dependence of the leading-to-trailing number ratio $R_{45}=N_4/N_5$ on the system parameters and evaluate the possible contribution of each parameter.

\begin{table}
 \hspace{-1 cm}
\centering
\begin{minipage}{8cm}
%\caption{\textbf{Similarly to Table \ref{model1}, but only for the standard setting of $\Delta\sigma_0=30^{\circ}$-$80^{\circ}$. We adopt different eccentricities ($e_J$) of Jupiter, and the first column corresponds to the same case of the real $e_J$ as that highlighted in bold in Table \ref{model1}.}}      % title of Table
\caption{The first row refers to the same case highlighted in bold in Table \ref{model1}, with Jupiter having the current eccentricity of $e_J=0.05$. In the two cases below, we adopt different $e_J$ by considering possible orbital excitation of Jupiter during the giant planet instability.}      % title of Table
\label{diff_eJ}
\begin{tabular}{c| c c c c c}        % centered columns (9 columns)
\hline                 % inserts double horizontal lines
$e_J$       &         $N_4$      &        $A_4$          &         $N_5$           &      $A_5$      & $R_{45}$                  \\

\hline
           
\textbf{0.05}            &          \textbf{312}         &      \textbf{15-54}        &         \textbf{209}              &        \textbf{18-56}       &        \textbf{1.50}           \\

0.1                            &          213                  &   14-45                    &        125       &   18-45        &         1.70                     \\

0.15                           &         136                   &   12-42                    &        55        &   17-38        &         2.47                      \\
\hline
\end{tabular}
\end{minipage}
\end{table}

\subsubsection{Jupiter's eccentricity and inclination}

Previously, we considered Jupiter having the eccentricity of $e_J=0.05$, i.e. the currently observed value. This small $e_J$ indicates that Jupiter's orbit is close to the circular case which is used in the theory developed for the distortion of phase space near the L4 and L5 points. In fact, during the giant planet instability, along with sudden changes in semi-major axis, Jupiter may also experience increases in $e_J$ \citep[e.g. Fig. 4 in][]{nesy13}. We noticed that $e_J$ could be excited to a value up to about 0.1. 

The eccentricity of Jupiter has influence on the shapes of the libration islands around the Lagrangian points \citep{li2021}, and thus may affect our results about the L4/L5 asymmetry caused by the fast migration of Jupiter. To validate our results presented earlier, for the case of the standard $\Delta\sigma_0$ setting in the control model, we carry out extra runs by choosing larger $e_J$ while keeping the other parameters unchanged. We consider a value of $e_J=0.1$ as mentioned above and additionally, a much more excited Jupiter with $e_J=0.15$ is provided for qualitatively understanding the effects of $e_J$. In Table \ref{diff_eJ}, for different $e_J$, we report the numbers of survived L4 and L5 Trojans, the corresponding number ratio and the final resonant amplitudes.

The first row in Table \ref{diff_eJ} refers to the case of $e_J=0.05$ which corresponds to the same results highlighted in bold in Table \ref{model1}. We find that the L4/L5 number ratio $R_{45}$ increases very fast with $e_J$. Comparing to $R_{45}=1.50$ at $e_J=0.05$, when $e_J$ is set to be 0.1 as found in \citet{nesy13}, the resulting $R_{45}$ acquires the larger value of 1.70. Bear in mind that, the longer Gyr evolution could further increase $R_{45}$ by a factor of 1.1, i.e. to $\sim1.9$. Then the unbiased L4/L5 asymmetry, with $R_{45}=1.6$, can be produced by given weaker requirements, e.g. a slower Jupiter's migration compared to the control model, or smaller $\Delta \sigma_0$ of Trojans compared to the standard setting. Moreover, at $e_J=0.15$, $R_{45}$ increases to a much larger value of 2.47. Therefore, the sudden increase of $e_J$ during the outward jump of Jupiter could be beneficial for explaining the observed L4/L5 asymmetry. Besides this major difference, a comparison with the case of $e_J=0.05$ raises two more differences. When the eccentricity of Jupiter increases: (1) both the numbers $N_4$ and $N_5$ of survived L4 and L5 Trojans decrease; (2) the maxima of the resonant amplitudes $A_4$ and $A_5$ become smaller, i.e. the test Trojans would survive on the orbits closer to the individual Lagrangian points. These results could be attributed to stronger perturbations of Jupiter on a more eccentric orbit.

Besides the eccentricity, the inclination $i_J$ of Jupiter could also affect the extent of the L4/L5 asymmetry in our jumping-Jupiter model. We note that, however, in the work about the giant planet instability by \citet{deie18}, they found that Jupiter's inclination was never too high, i.e. $i_J<5^{\circ}$. Actually, in our numerical simulations Jupiter has an inclination of $i_J=1.3^{\circ}$ with respect to the reference plane (i.e. the ecliptic plane) and this value is comparable with the said upper limit of $i_J$.  We also think that a small change in Jupiter's inclination would not make an essential different in the current results.

%Second, since we will consider Trojans with inclinations $i$ as high as tens of degrees, this will be equivalent to having a Jupiter with higher $i_J$ and Trojans with low $i$, because in both cases the mutual inclination between Jupiter and the Trojan population can have similar values. Thus the effect of Jupiter's inclination can be included in the case of inclined Trojans, as discussed below.

% even if $i_J$ was excited to several degrees ($>2^{\circ}$, not very likely), the relative inclinations between Jupiter and its Trojans would still be similar the large values of $i$. 

\subsubsection{Inclination distribution of Trojans}

Analogous to the investigation of different eccentricities of Jupiter reported above, we can evaluate the influence of the orbital parameters of Jupiter Trojans on their leading-to-trailing number ratio $R_{45}$. We may need to consider the Trojan swarms having eccentricity and inclination distributions according to the current observation; this is because they should have been captured with such orbital distributions. In previous simulations, we adopted test Trojans with initial eccentricities $e=0$-0.3 (which covers the observational range), while their initial inclinations were taken to be $i\sim0$ as in the coplanar 3-body model used for theoretically exhibiting the distortions of the L4 and L5 islands.

As we have shown, the outward jump of Jupiter can well resolve the L4/L5 asymmetry for Trojans with $i\sim0$. Then it is of great interest to investigate whether this mechanism could be applied to more inclined Trojans. By simply analyzing the latest orbital data of Jupiter Trojans recorded in the MPC, we notice that the real Jupiter Trojans present inclinations up to about $i=40^{\circ}$-50$^{\circ}$, while a vast majority ($\sim95$\%) of them have relatively smaller $i$ of $<30^{\circ}$.

For the test Trojans with standard $\Delta\sigma_0$ setting, their initial orbital parameters are the same as those in the case of $i\sim0$ (i.e. the bold case in Table \ref{model1}), but the inclinations are set to three other representative values of $i=10^{\circ}$, $20^{\circ}$ and $30^{\circ}$. Then we re-simulate the evolution of these inclined Trojans in the control model. The results are listed in Table \ref{diff_iJT}. For reference, the case of $i\sim0$ is also presented in the first row. One would immediately observe that, all the cases with different $i$ can result in noticeable L4/L5 asymmetry. Although the number ratio $R_{45}$ becomes smaller at higher $i$, the drop-off is quite slow, only from $\sim1.5$ ($i\le10^{\circ}$) to 1.3 ($i=30^{\circ}$). If we simply calculate the global asymmetry by averaging $R_{45}$ over different $i$, the obtained leading-to-trailing number ratio is 1.43. By including the effect of the later long-term evolution, the value of $R_{45}=1.43\times1.1\approx1.57$ can also account for the unbiased observation (i.e. $R_{45}=1.6$).

Nevertheless, in Table \ref{diff_iJT} for surviving Trojans, the number ratio $R_{45}$ does not decrease monotonically with increasing inclination $i$. The turning point is at $i=10^{\circ}$, where the corresponding $R_{45}=1.54$ is about the same as that in the case of $i\sim0$ ($R_{45}=1.50$). We speculate that the cause may be the not very high $i$. Naturally, we check an additional case of $i=5^{\circ}$ and we find that the resultant $R_{45}=1.49$ is still similar in value. Accordingly, we infer that the inclinations of Jupiter Trojans would have an impact on the L4/L5 asymmetry, but within a restricted range, and hence our main results would not change. A more detailed discussion about the real inclination distribution will be made in the next section.

\begin{table}
 \hspace{-1 cm}
\centering
\begin{minipage}{8cm}
%\caption{\textbf{Similarly to Table \ref{model1}, but only for the standard setting of $\Delta\sigma_0=30^{\circ}$-$80^{\circ}$. We adopt different inclinations ($i$) of Jupiter Trojans, and the first column corresponds to the same case of $i\sim0$ as that highlighted in bold in Table \ref{model1}.}}      % title of Table
\caption{The first row refers to the same case highlighted in bold in Table \ref{model1}, for Jupiter Trojans with inclinations $i\sim0$. Below are the three typical inclined cases of $i=10^{\circ}$, $20^{\circ}$ and $30^{\circ}$.}
\label{diff_iJT}
\begin{tabular}{c| c c c c c}        % centered columns (9 columns)
\hline                 % inserts double horizontal lines
$i$       &         $N_4$      &        $A_4$          &         $N_5$           &      $A_5$      & $R_{45}$                  \\

\hline
           
$\sim$\textbf{0}            &          \textbf{312}         &      \textbf{15-54}        &         \textbf{209}              &        \textbf{18-56}       &        \textbf{1.50}           \\

%$5^{\circ}$                           &          318                  &   15-51                    &        213       &   18-55       &         1.49                    \\

$10^{\circ}$                          &          306                  &   14-52                    &        199       &   18-56        &         1.54                    \\

$20^{\circ}$                          &         328                   &   16-53                  &        238        &   19-60        &         1.38                      \\

$30^{\circ}$                          &         349                   &   17-53                    &        269       &   17-63        &         1.30                      \\
\hline
\end{tabular}
\end{minipage}
\end{table}

\section{Conclusions and discussion}

The number asymmetry of Jupiter Trojans, i.e. the leading (L4) swarm hosting a larger population than the  trailing (L5) swarm, is a long-lasting problem. This distinctive feature can be very important to understand the early evolution of the Solar System. Actually, the observed L4-to-L5 count ratio $R_{45}=N_4/N_5$ changes as more and more Trojans are discovered, and currently the value of $R_{45}$ is about 1.8.  Numerous studies have been carried out investigating how this number difference could be affected by observational biases. \citet{szab07} estimated that the unbiased $R_{45}$ is about 1.6, which is used to constrain the migration of Jupiter and the evolution of the Trojans in this paper.

After the planetesimals were captured into the Trojan population, the L4 or L5 swarm could initially contain a similar number of objects. Here, we consider the subsequent evolution of the system when Jupiter happened to experience an outward jump and the librational spaces around the L4 and L5 points were distorted, possibly leading to different survival rates of the leading and trailing Trojans. The key factor in this mechanism is the migration speed of Jupiter, which determines the shapes of the L4 and L5 librational regions.

The Jupiter jump could be caused by scattering encounters with an extra ice giant. We adopt a migration amplitude $\Delta a_J=0.5$ AU and an e-folding time $\tau=1000$ yr, giving a migration speed of $\dot{a}_J=\Delta a_J/(10~\tau/3)=1.5\times10^{-4}$ AU/yr. This speed value is supported by previous works \citep[e.g.][]{nesy13}, and the corresponding outward migration of Jupiter can induce the contraction of the L4 region and the expansion of the L5 region \citep{sica03}. 
From numerical simulations, we find that the L4 Trojans have their resonant amplitudes decreased and become more stable, while the L5 Trojans have their resonant amplitudes increased and some of them finally get ejected (see Fig. \ref{RealJTs}). As a result, this jumping-Jupiter model could induce more stable L4 Trojans and less stable L5 Trojans, and accordingly we expect to get the increase of $R_{45}$. 

Then we constructed numerical experiments to simulate the evolution of the L4 and L5 swarms starting with symmetric resonant angles of $60^{\circ}+\Delta \sigma_0$ and $-60^{\circ}-\Delta \sigma_0$ respectively, where $\Delta \sigma_0$ is a certain range. Given the same number of bodies for these two Trojan swarms, we find that for various choices of $\Delta \sigma_0$, the number ratios $R_{45}$ of the survived Trojans are always larger than 1. In particular, when we adopt a standard $\Delta\sigma_0$ setting of $30^{\circ}$-$80^{\circ}$, the resulting $R_{45}$ can be as large as 1.5 at the end of the 1 Myr integration. During the longer Gyr evolution, $R_{45}$ could further increase and finally reach a value of $\sim1.6$, accounting for the unbiased number asymmetry. In addition, for the standard setting of $\Delta \sigma_0$, we have investigated a variety of Jupiter migration rates within a plausible range. The results verified that the faster the Jupiter's outward migration is, there are more L4 survived Trojans than L5 ones. Thus, $R_{45}$ can achieve an even larger value. More interestingly, when the migration of Jupiter is fast enough, the L4 librational islands are significantly diminished, consequently the local Trojan become very stable on their orbits with rather small resonant amplitudes and nearly all of them can survive even if their initial displacements from the L4 point are quite large. 

To strengthen the robustness of our conclusions in the jumping-Jupiter model, we carried out further investigations to consider the effects of other system parameters on the L4/L5 asymmetry. We showed that: (1) during the instability of giant planets, the eccentricity $e_J$ of Jupiter could be excited to a value higher than the current eccentricity of 0.05, e.g. about 0.1, as found in \citet{nesy13}. The increase of $e_J$ can induce a larger value of $R_{45}$, and thus make the L4/L5 asymmetry easier to be explained. (2) Jupiter's inclination $i_J$ could experience an excitation during the early evolution, while \citet{deie18} gave an upper limit of $i_J<5^{\circ}$. This $i_J$ is comparable to the current value that we used in the control model, indicating that $i_J$ should not play an important role. (3) As the inclinations of Trojans increase, the leading-to-trailing number ratio $R_{45}$ would drop, but very slowly. At this point, we justified that the overall asymmetry for Trojans with a wide range of inclinations can also be well consistent with the observations.

In summary, the mechanism of jumping-Jupiter and distortion of the L4/L5 tadpole orbits could potentially explain the unbiased number asymmetry of $R_{45}\sim1.6$ for the known Jupiter Trojans. We would also like to mention that, even if some number asymmetry of Jupiter Trojans was pre-existing, the present work would still be valid and further increase the value of $R_{45}$. Nevertheless, there exists a problem that, in the cases that we obtain $R_{45}\sim1.6$, the simulated Trojans generally have quite larger resonant amplitudes compared to the real ones. This problem is to be tackled with some dissipation of energy, e.g. by mutual collisions, as we addressed it in the main text.

In this paper, we assume that the Jupiter Trojans were captured at the stage of the giant planet instability as in \citet{nesy13}. During the later jumps, once moving inwards, Jupiter lost more L4 Trojans, and when moving outwards, it lost more L5 Trojans. Therefore, if the numbers of inward and outward jumps of Jupiter eventually balanced out, the numbers of the Trojans around L4 and L5 would be nearly the same. This may explain why \citet{nesy13} did not find the same results as ours regarding the L4/L5 asymmetry. In fact, our work provides an evidence that the number of Jupiter's jumps did not balance out, e.g. Jupiter performed $N+1$ outward jumps compared to $N$ inward jumps.

Differently from \citet{nesy13} where the last jump was often inward, there was an extra large outward jump in our model. This outward jump can cause the L4/L5 asymmetry of Jupiter Trojans. Then, the fifth outer planet would get ejected and Jupiter would enter in its tail migration mode (i.e. planetesimal-driven migration) which is mostly inwards, smooth and at low speed \citep{nesy12, deie17}. However, there are several caveats that we need to acknowledge for the interpretation of having Jupiter performing the last jump outwards. The first is the implication that Jupiter was not the celestial body responsible for ejecting the fifth outer planet. That is because, from the theory of angular momentum exchange, Jupiter necessarily needs to jump inwards once it ejects a planet. An alternative way may be that this planet simply collided with the Sun instead of being ejected. Another issue is that, before the last outward jump with the considered $\Delta a_J$ range, we need to assume that Jupiter was at $a_J < 4.9$ AU, which would put Jupiter much closer to the main belt asteroids. Otherwise, the tail planetesimal-driven (inward) migration of Jupiter has to be longer than the usual distance of $\sim$0.2 AU at maximum. The lengthier the tail migration was, the larger the amount of mass encountering Jupiter should be, thus the primordial planetesimal disk would be more massive. As a result of too much disk mass, the eccentricity of Jupiter would be too low at the end of the long tail migration, compared to the current value of $\sim0.05$ \citep{nesy12, deie17}. The possible solutions are to be derived and in addition, a more robust evolutionary pathway of Jupiter is projected for our future works.

\subsection{Future work}

%In this paper, our hypothetical models are to test the effect of a jumping-Jupiter on the number asymmetry of the Jupiter Trojans. However, these models are quite rough, and we should carry out more detailed study by considering the said dynamical aspects of the planetary system evolution. Future works would help in establishing a more realistic model. 

In this paper, our hypothetical model is to test a possible explanation to the number asymmetry of the Jupiter Trojans. However, this model seems quite rough, and future works would help in establishing a more delicate scenario. Some dynamical aspects of the planetary system evolution that are worthwhile to consider are discussed below:

(1) According to \citet{nesy13}, we used a representative value of Jupiter's migration timescale $\Delta t$, which is of the order of $10^4$ yr. Actually, the jump of Jupiter can be much faster if the planet-planet close encounter is deeper. Bearing in mind that, when Jupiter's migration speed $\dot{a}_J$ exceeds a critical value of $\dot{a}_J^{crit}=3.8\times10^{-3}$ AU/yr, L4 will merge with L3 and the surrounding tadpole islands will disappear \citep{sica03, ogil06}. Thus, the evolution of the L4 Trojan swarm will be much different and the extent of the L4/L5 number asymmetry could be further enhanced.

%we used a representative value of the migration timescale $\Delta t\sim10^4$ yr from \citet{nesy13}, the jump of Jupiter would be much faster if the planet-planet close encounter is deeper. Bearing in mind that, when Jupiter's migration speed $\dot{a}_J$ exceeds the value of $\dot{a}_J^{crit}=3.8\times10^{-3}$ AU/yr, the L4 point would disappear \citep{sica03, ogil06}. Thus, it is very interesting to see what would happen to the L4 Trojan swarm if $\dot{a}_J>\dot{a}_J^{crit}$, as we will publish in a follow up paper.

(2) Regarding the cumulating loss of Trojans over several inward and outward jumps of Jupiter, an interesting question is that, does the order of jumps matter or not? As mentioned previously, for $N+1$ outward jumps of Jupiter compared to $N$ inward jumps during the giant planet instability, we simply assumed that the last jump was an outward one. More sophisticated and detailed investigation is to build a sequence of jumps of Jupiter, inward and outward, and cumulate the number losses for both the L4 and L5 Trojans, assuming that the same quantity of new Jupiter Trojans would be captured around each Lagrangian point after every jump. If finally, different combinations of $N+1$ outward and $N$ inward jumps of Jupiter could always generate the current L4/L5 asymmetry, this result would be even robust and may potentially constrain Jupiter's evolution at the stage of the giant planet instability.

(3) In order to simplify our model, we did not take into account the effects of Saturn, Uranus and Neptune in our simulations. This is reasonable because it is much straightforward to compare the numerical results with the theoretical conjecture. However, these three planets do exist and may have influences on the motions of Jupiter Trojans, especially the closest planet Saturn. By performing numerical simulations, \citet{marz02} and \citet{marz03} have shown that, the direct perturbation by Saturn is a source of instability for Jupiter Trojans on a long timescale of $10^7$-$10^8$ yr. Since in our model the L4/L5 number asymmetry is mainly generated during the short period of Jupiter jumping, within a few thousand years, we suppose that such secular perturbations would not affect the key results of this paper. As a matter of fact, due to the additional influence of Saturn, the L4 Trojan swarm could be more stable than the L5 swarm \citep{frei06}, i.e. further enhancing the L4/L5 number asymmetry that we obtained. Another consideration of excluding the other three giant planets besides Jupiter is that, at the stage of Jupiter jumping in the early time, the architecture of the outer Solar System remains quite uncertain. Therefore the arbitrarily designed model (e.g. the orbit of Saturn) may be leading to unreliable results.

% Saturn was neglected to reduce the numerical computations, it has indirect effect on the motions of Jupiter Trojans, and is not expected to influence the number asymmetry considered here. at the early stage of Jupiter jumping, the architecture of outer solar system could be much different, thus we can not simple include the other three giant planets
% Our hypothetical model is to test the effect of a jumping-Jupiter on the number asymmetry. We did not invoke Saturn in the jumping-Jupiter model, its perturbations may affect the long-term stability of Trojans, but not the short period of Jupiter jumping within thousands years, 
% in the current orbits, the L4 Trojan swarm seems to be more stable than the L5 swarm due to the influence of Saturn \citep{frei06}

(4) We have investigated whether the jumping-Jupiter model could produce the L4/L5 number asymmetry of the Jupiter Trojans, at the level of $R_{45}\sim1.6$ for the unbiased observation. Since the intrinsic inclination distribution of the Jupiter Trojans to date has not been well-modelled \citep{park15}, we first considered Jupiter and its Trojan population having nearly coplanar orbits and then the relative inclinations were adopted to have some particular values of up to $30^{\circ}$. \citet{slyu13} shows that with respect to the inclination, the L4 number distribution is essentially different from the L5 one, and the confidence level is as high as 99\%.  It is also important to point out that, at small inclinations $<5^{\circ}$, there is still a greater number of Trojans belonging to the L4 group. As we found before, our jumping-Jupiter model constructed to explain the number difference between the L4 and L5 Trojans could partly rely on their inclinations, but not strongly. Our next step is to develop and generalise this model for the real Trojan population, which has a broad but not well determined inclination distribution.

%(3) We have investigated whether the jumping-Jupiter model could produce the L4/L5 number asymmetry of the Jupiter Trojans, at the level of $R_{45}\sim1.6$ for the unbiased observation. Since the intrinsic inclination distribution of the Jupiter Trojans to date has not been well-modelled \citep{park15}, we first considered the planet-Trojans system having nearly coplanar orbital geometry. \citet{slyu13} shows that, with respect to the inclination, the L4 number distribution is essentially different from the L5 one, and the confidence level is as high as 99\%.  It is also important to point out that, at small inclinations $<5^{\circ}$, there is still a greater number of Trojans belonging to the L4 group. We suppose that our jumping-Jupiter model constructed to explain the number difference between the L4 and L5 Trojans could weakly rely on their inclinations. Our next step is to develop and generalise this model for the realistic Trojan population, which has a broad inclination distribution up to $55^{\circ}$.

Finally, we would like to mention that, the observational features of Jupiter Trojans can help us further improve our jumping-Jupiter model. For example, based on the latest survey results, \citet{ueha22} suggested that the L4/L5 number asymmetry of Jupiter Trojans with diameters $\ge 2$ km could be about 1.4, which is lower than the number ratio $R_{45}$ of 1.6 that we considered in this paper. Therefore, the migration speed of Jupiter needs to be adjusted accordingly. Besides, the color and size distributions of Jupiter Trojans are similar to those of Kuiper belt objects \citep{Fras14, wong16}, both these features place constrains on the primordial progenitor population for Jupiter Trojans. Then we have to deal with more comprehensive initial conditions of our test Trojans.

%\begin{figure*}
%  \centering
% \begin{minipage}[c]{1\textwidth}
%  \vspace{0 cm}
% \includegraphics[width=9cm]{Fig1a.jpg}
%  \includegraphics[width=9cm]{Fig1c.jpg}
%  \end{minipage}
%  \begin{minipage}[c]{1\textwidth}
%  \vspace{-0.25 cm}
%  \includegraphics[width=9cm]{Fig1b.jpg}
%  \includegraphics[width=9cm]{Fig1d.jpg}
%  \end{minipage}
%  \vspace{0 cm}
%  \caption{Distribution }
% \label{Obs}
% \end{figure*}  

% \begin{figure*}
%  \centering
%  \begin{minipage}[c]{1\textwidth}
%  \vspace{0 cm}
%  \includegraphics[width=9cm]{Fig2a.jpg}
%  \includegraphics[width=9cm]{Fig2b.jpg}
%  \end{minipage}
%  \vspace{0 cm}
%  \caption{As in the up}
% \label{ObsMyr}
% \end{figure*}  

%\begin{figure}
% \hspace{0 cm}
%  \includegraphics[width=8.5cm]{Fig6.jpg}
%  \caption{For }
%  \label{time}
%\end{figure}

%\begin{equation}
%\mu_9\approx +n\frac{1}{4}\frac{m_9}{m_{\odot}}{\alpha_9}{\bar{\alpha}_9}b_{3/2}^{(1)}({\alpha_9})\sin i_9,
%\label{mu9}
%\end{equation}

%\begin{eqnarray}
%  &&q_0=-\sum_{i=1}^{9}\frac{\mu_i}{B-f_i}\cos(f_i t + \gamma_i),\nonumber\\
%  &&p_0=-\sum_{i=1}^{9}\frac{\mu_i}{B-f_i}\sin(f_i t + \gamma_i),
%\label{forced}
%\end{eqnarray}

\begin{acknowledgements}
    
This work was supported by the National Natural Science Foundation of China (Nos. 11973027, 11933001), and National Key R\&D Program of China (2019YFA0706601). And part of this work was also supported by a Grant-in-Aid for Scientific Research (20H04617). We would also like to express our sincere thanks to the referee for his valuable comments, which helped to improve the quality of this paper.
%We would also like to express our sincere thanks to the anonymous referee for the valuable comments.
      
\end{acknowledgements}


\end{document}